\documentclass[fleqn,usenatbib]{mnras}

\usepackage{newtxtext,newtxmath}

\usepackage[T1]{fontenc}

\DeclareRobustCommand{\VAN}[3]{#2}
\let\VANthebibliography\thebibliography
\def\thebibliography{\DeclareRobustCommand{\VAN}[3]{##3}\VANthebibliography}

\def\usableSNe{1577}
\def\FoundationSNe{123}
\def\LOWZSNe{76}
\def\LOWZFOUND{199}
\def\nsimSNe{1450}

\def\SOMPZbias{0.023}
\def\SOMPZscatter{0.071}
\def\SOMPZoutlier{0.337}
\def\BPZbias{-0.002}
\def\BPZscatter{0.041}
\def\BPZoutlier{0.191}
\def\DNFbias{0.000}
\def\DNFscatter{0.020}
\def\DNFoutlier{0.106}

\def\SNDNFbias{-0.003}
\def\SNDNFscatter{0.019}

\def\simSPEChscatter{0.152}
\def\simIASPEChscatter{0.143}
\def\simDNFhscatter{0.176}

\def\dataSPEChscatter{0.167}
\def\dataIASPEChscatter{0.160}

\def\simSOMPZPZdw{$-0.007 \pm 0.003$}
\def\simSOMPZGdw{$-0.006 \pm 0.002$}

\def\simDNFdw{$-0.003 \pm 0.002$}

\def\dataSOMPZPZdw{$-0.019$}
\def\dataSOMPZGdw{$-0.002$}

\def\dataDNFdw{$-0.017$}

\def\simCCdw{$-0.0008$}

\usepackage{graphicx}	
\usepackage{amsmath}	

\newcommand{\photoz}{photo-$z$}
\newcommand{\specz}{spec-$z$}

\usepackage{eso-pic}

\AddToShipoutPictureBG*{%
  \AtPageUpperLeft{%
    \hspace{0.73\paperwidth}%
    \raisebox{-5.5\baselineskip}{%
      \makebox[0pt][l]{\textnormal{DES-2023-0819}}
}}}%

\AddToShipoutPictureBG*{%
  \AtPageUpperLeft{%
    \hspace{0.73\paperwidth}%
    \raisebox{-4.5\baselineskip}{%
      \makebox[0pt][l]{\textnormal{FERMILAB-PUB-24-0360-PPD}}
}}}%

\title[DES-SN5YR photo-$z$]{Evaluating Cosmological Biases using Photometric Redshifts for Type Ia Supernova Cosmology with the Dark Energy Survey Supernova Program}

\author[Chen et al.]{
\parbox{\textwidth}{
\Large
R.~Chen,$^{1}$
D.~Scolnic,$^{1}$
M.~Vincenzi,$^{1,2,3}$
E.~S.~Rykoff,$^{4,5}$
J.~Myles,$^{6}$
R.~Kessler,$^{7,8}$
B.~Popovic,$^{1,9}$
M.~Sako,$^{10}$
M.~Smith,$^{11}$
P.~Armstrong,$^{12}$
D.~Brout,$^{13,14}$
T.~M.~Davis,$^{15}$
L.~Galbany,$^{16,17}$
J.~Lee,$^{10}$
C.~Lidman,$^{18,12}$
A.~M\"oller,$^{19}$
B.~O.~S\'anchez,$^{20,1}$
M.~Sullivan,$^{11}$
H.~Qu,$^{10}$
P.~Wiseman,$^{11}$
T.~M.~C.~Abbott,$^{21}$
M.~Aguena,$^{22}$
S.~Allam,$^{23}$
O.~Alves,$^{24}$
F.~Andrade-Oliveira,$^{25}$
J.~Annis,$^{23}$
D.~Bacon,$^{26}$
D.~Brooks,$^{27}$
A.~Carnero~Rosell,$^{28,22}$
J.~Carretero,$^{29}$
A.~Choi,$^{30}$
C.~Conselice,$^{31,32}$
L.~N.~da Costa,$^{22}$
M.~E.~S.~Pereira,$^{33}$
H.~T.~Diehl,$^{23}$
P.~Doel,$^{27}$
S.~Everett,$^{34}$
I.~Ferrero,$^{35}$
B.~Flaugher,$^{23}$
J.~Frieman,$^{23,8}$
J.~Garc\'ia-Bellido,$^{36}$
M.~Gatti,$^{10}$
E.~Gaztanaga,$^{16,26,17}$
G.~Giannini,$^{29,8}$
D.~Gruen,$^{37}$
R.~A.~Gruendl,$^{38,39}$
G.~Gutierrez,$^{23}$
K.~Herner,$^{23}$
S.~R.~Hinton,$^{15}$
D.~L.~Hollowood,$^{40}$
K.~Honscheid,$^{41,42}$
D.~Huterer,$^{24}$
D.~J.~James,$^{43}$
K.~Kuehn,$^{44,45}$
M.~Lima,$^{46,22}$
J.~L.~Marshall,$^{47}$
J. Mena-Fern{\'a}ndez,$^{48}$
F.~Menanteau,$^{38,39}$
R.~Miquel,$^{49,29}$
R.~L.~C.~Ogando,$^{50}$
A.~Palmese,$^{51}$
A.~Pieres,$^{22,50}$
A.~A.~Plazas~Malag\'on,$^{4,5}$
A.~Roodman,$^{4,5}$
S.~Samuroff,$^{52}$
E.~Sanchez,$^{53}$
D.~Sanchez Cid,$^{53}$
I.~Sevilla-Noarbe,$^{53}$
E.~Suchyta,$^{54}$
M.~E.~C.~Swanson,$^{38}$
G.~Tarle,$^{24}$
C.~To,$^{41}$
D.~L.~Tucker,$^{23}$
V.~Vikram,$^{10}$
N.~Weaverdyck,$^{55,56}$
and J.~Weller$^{57,58}$
\begin{center} (DES Collaboration) \end{center}
}
\vspace{0.4cm}
\\
\parbox{\textwidth}{
\it Affiliations shown at end of paper
}
}

\date{Accepted XXX. Received YYY; in original form ZZZ}

\pubyear{\the\year{}}

\begin{document}
\label{firstpage}
\pagerange{\pageref{firstpage}--\pageref{lastpage}}
\maketitle

\begin{abstract}
Cosmological analyses with Type Ia Supernovae (SNe Ia) have traditionally been reliant on spectroscopy for both classifying the type of supernova and obtaining reliable redshifts to measure the distance-redshift relation. While obtaining a host-galaxy spectroscopic redshift for most SNe is feasible for small-area transient surveys, it will be too resource intensive for upcoming large-area surveys such as the Vera Rubin Observatory Legacy Survey of Space and Time, which will observe on the order of millions of SNe. Here we use data from the Dark Energy Survey (DES) to address this problem with photometric redshifts (\photoz) inferred directly from the SN light-curve in combination with Gaussian and full $p(z)$ priors from host-galaxy \photoz\ estimates. Using the DES 5-year photometrically-classified SN sample, we consider several \photoz\ algorithms as host-galaxy \photoz\ priors, including the Self-Organizing Map redshifts (SOMPZ), Bayesian Photometric Redshifts (BPZ), and Directional-Neighbourhood Fitting (DNF) redshift estimates employed in the DES 3x2 point analyses. With detailed catalog-level simulations of the DES 5-year sample, we find that the simulated $w$ can be recovered within $\pm0.02$ when using SN+SOMPZ or DNF prior \photoz, smaller than the average statistical uncertainty for these samples of 0.03. With data, we obtain biases in $w$ consistent with simulations within $\sim1\sigma$ for three of the five \photoz\ variants. We further evaluate how \photoz\ systematics interplay with photometric classification and find classification introduces a subdominant systematic component. This work lays the foundation for next-generation fully photometric SNe Ia cosmological analyses.
\end{abstract}




\section{Introduction}
Type Ia Supernovae (SNe Ia) are standardizable candles used as a key cosmological probe to measure the distance-redshift relation and understand the nature of dark energy. In order to precisely constrain the dark energy equation-of-state parameter \textit{w}, the largest SN surveys and compilations to date such as Pantheon+ \citep{PantheonPlus} have relied on access to spectroscopy of the live SN to confirm the type of the SN, as well as of the host galaxy to obtain an accurate redshift. Recent cosmological analyses such as the Dark Energy Survey 5-year SN analysis (DES-SN5YR; \citealp{Vincenzi24}, \citealp{DES5YR}) have shown that a transition to photometric classification immensely boosts our statistical constraining power without introducing significant systematic uncertainty. Upcoming surveys such as the Vera Rubin Observatory Legacy Survey of Space and Time (LSST; \citealp{Ivezic19}) and the Nancy Grace Roman Space Telescope High-Latitude Time-Domain Survey (henceforth Roman; \citealp{Spergel15}, \citealp{Hounsell18}, \citealp{Rose21}) are poised to observe multiple orders of magnitude more supernovae than our current largest compilations. As it becomes impossible to spectroscopically follow-up every SN or its host galaxy, photometric classification will increase the number of cosmologically suitable SNe. However, cosmological analyses with these next-generation datasets will still be limited if they rely on spectroscopic redshifts (\specz) from host galaxies. Therefore, efforts to develop the usability of photometric redshifts (\photoz) for Ia cosmology, either from the SN, the host galaxy, or a combination of the two, will be critical to optimize the potential of Stage IV Dark Energy experiments. 

Photometric redshift estimation is an area of active research, as other key cosmological probe analyses already require redshifts for hundreds of millions of galaxies for which it is impossible to obtain spectroscopic redshifts. In weak gravitational lensing, these efforts are focused on characterizing the redshift distributions of source and lens galaxy samples in several tomographic bins, i.e.\ $n(z)$, to extreme precision, for example using Self-Organizing Map \photoz\ (SOMPZ; \citealp{Buchs19}, \citealp{Myles21}). Other photometric redshift estimates are primarily concerned with the performance of individual galaxy redshift estimates. Several data-driven approaches make use of machine-learning methods to learn the mapping between colors and redshift (e.g. \citealp{TPZ}, \citealp{ANNz2}, \citealp{DNF}), while others rely on templates of galaxy SEDs to find the best fit redshift (e.g. \citealp{BPZ}, \citealp{Feldmann06}, \citealp{Brammer08}). Each method is sensitive to different systematics that affect the precision and accuracy of the final redshift estimates for different galaxy populations to varying degrees. Each \photoz\ estimate can be given as a fully descriptive probability distribution function (PDF) but are frequently simplified to a point estimate (mean or mode) and a Gaussian uncertainty. 

Previous analyses have illustrated the potential for using photometric redshifts for Type Ia Supernova cosmology. \citet{Chen22} performed an analysis using a subset of the DES-SN5YR sample containing SNe with redMaGiC (a selection of Luminous Red Galaxies; \citealp{Rozo16}) host galaxies. These galaxies have particularly well-constrained photometric redshifts, with $\frac{\sigma_{z}}{1+z} < 0.02$, but comprise only $\sim$6\% of the DES-SN5YR sample. Using this sample of $\sim$125 SNe, they found that using the redMaGiC host galaxy \photoz\ directly in place of the spectroscopic redshift results in a $w$-shift of $\sim0.005$, and this result was validated with consistent results from detailed catalog-level simulations. However, this analysis did not model the effects of core-collapse contamination. redMaGiC host galaxies are expected to have low to zero rates of core-collapse supernovae, as they are old, passive galaxies, compared to the star-forming galaxies that core-collapse progenitors occur in \citep{Irani22}. This analysis was further simplified by restricting the sample host galaxies to a single type, reducing the need for modeling dependencies between the SN and host-galaxy properties. \citet{RuhlmannKleider22} present an analysis using a combined sample of the spectroscopic Joint Light-curve Analysis (JLA) SN Ia sample and the 3-year photometric SuperNova Legacy Survey (SNLS) SN Ia sample. Using a mixture of spectroscopic and photometric redshifts, they found that while a naive analysis with photometric redshifts and contamination leads to biased matter density ($\Omega_{\textrm{M}}$) for a Flat-$\Lambda$CDM model, the bias can be corrected with an appropriately modeled magnitude bias correction computed for selection effects.

An alternative to directly using the host-galaxy \photoz\ is to infer the \photoz\ from the supernova light-curve. This can be done by extending the SALT (Spectral Adaptive Light-curve Template; \citealp{Guy10}) SN light-curve model fitting framework such that the SN redshift is floated, rather than fixed in the fit. These estimates are greatly improved by optionally including a prior on the redshift from a host-galaxy \photoz\ estimate (SN+host \photoz: \citealp{Kessler10a, Palanque10, Dai18}). This \photoz\ approach has two unique benefits: i) it combines information from two independent \photoz\ estimates, whereas other probes rely on a single galaxy \photoz\ and are thus more vulnerable to catastrophic outliers, and ii) the redshift covariance is propagated to the other SALT fitted parameters. This inferred \photoz\ method was applied to the extended DES-SN5YR sample classified without redshifts from \citet{Moller24}. They used host \photoz\ priors only when available and evaluated the biases of this mixed sample. Using simulations they found structured offsets in redshift and SALT2 parameter estimation that can be minimized by binning. However, they did not evaluate the impact on cosmology. \citet{Mitra23} presented an analysis using a set of mock-data simulated LSST SNe from the Deep Drilling Fields and found that supplementing the subset of SNe with \specz\ with SN+host \photoz\ increases the $w_{0}w_{a}$ figure-of-merit by 50\%. In particular, this helps extend the upper limit of the redshift range of usable SNe, as obtaining spectroscopy becomes less feasible for fainter sources. While this LSST analysis included a statistical+systematic covariance matrix including calibration uncertainties, it did not include non-Ia contaminants or intrinsic scatter systematics.

In this work, we present an evaluation of cosmological biases for the Flat-$w$CDM model using data from the DES-SN5YR photometrically classified SN sample and SN+host galaxy photometric redshifts. We also present an implementation of SN+host \photoz\ that utilizes a full \photoz\ estimate PDF from the host galaxy rather than a point estimate with Gaussian uncertainty. To better evaluate our methodology, we focus solely on the subset of DES-SN5YR SNe with host spectroscopic redshifts and thus do not evaluate the statistical impact of a 5YR sample that includes events without a spectroscopic redshift. For the anchoring low-$z$ sample we use spectroscopic redshifts only.

The outline of the paper is as follows. In Section 2, we detail the DES-SN5YR photometric supernova sample, the anchoring low-$z$ sample, and the associated host-galaxy photometric redshift estimates and properties used for the analysis. In Section 3, we detail the catalog-level simulations used to validate the analysis and compute bias corrections for cosmology. In Section 4, we describe the analysis framework and cosmological parameter inference formalism. In Section 5, we present the results of the analysis for simulations and data. Finally, we discuss future prospects in Section 6 and conclude in Section 7.

\section{Data}
\label{sec:data}

\subsection{The Dark Energy Survey}
We use data from the Dark Energy Survey Supernova Program, which observed for five seasons with the Dark Energy Camera (DECam; \citealp{Flaugher15}) on the 4 meter Blanco telescope at Cerro Tololo Inter-American Observatory in Chile. DES consists of two programs: a wide field survey optimized for weak gravitational lensing, galaxy clustering, and galaxy cluster cosmology, and a time-domain survey primarily for SN cosmology. The ten SN fields lie within the footprint of the wide field but were observed with a much higher cadence to detect transients. For both programs, raw images were pre-processed by the DES Data Management team \citep{DESDM} to produce calibrated images and catalogs.

\subsubsection{DES SN Data}
The DES SN fields consist of ten 2.7 square degree fields, which were observed in the \textit{griz} filters at a cadence of $\sim7$ days. Of these ten fields, two `deep' fields (X3,C3) were observed to a single-visit depth of 24.5 mag in each band, and eight `shallow' fields (X1,X2,C1,C2,E1,E2,S1,S2) were observed to a depth of 23.5 mag. To identify transient candidates, images were processed through DIFFIMG \citep{Kessler15}, the DES difference imaging pipeline. Imaging artefacts were further removed using the AUTOSCAN algorithm \citep{Goldstein15}. A candidate light-curve was stored and updated if there were at least two detections within 1 arcsec, separated by at least 1 but not more than 30 days \citep{Kessler15}. 

While DIFFIMG delivered real-time SN photometry with $\sim2\%$ precision, our DES-SN5YR cosmology analysis includes a more precise Scene Modeling Photometry (SMP; \citealp{Holtzman08}, \citealp{Astier14}, \citealp{Brout19b}), simultaneously modeling the time-varying SN flux and the static host-galaxy flux. With SMP, 19,706 transient candidates were analyzed. \citet{Sanchez24} provides further details on the DES-SN5YR photometry. Corrections for Differential Chromatic Refraction (DCR; \citealp{Filippenko82}), which causes wavelength-dependent effects on the SN PSF shapes and positions, were computed for the DES-SN5YR sample \citep{LeeAcevedo22} but are not included here, as they are a subdominant effect on distance measurements. 

Host galaxies are assigned based on the Directional Light Radius method (DLR; \citealp{Sullivan06}, \citealp{Gupta16}) using a deep host galaxy library built from coadds \citep{Wiseman20, Qu23a}. While the DES-SN3YR cosmological analysis \citep{DES3YR} relied on a subset of spectroscopically typed SNe ($\sim200$), the DES-SN5YR cosmological analysis \citep{DES5YR} uses a sample of photometrically classified SNe ($\sim1600$). These SNe comprise the DES-SN5YR cosmology sample and are selected using only their light-curves and host-galaxy spectroscopic redshifts using SuperNNova \citep{Moller20, Moller22}. For host galaxy spectroscopic redshifts, DES relied on a partner program (OzDES) on the Anglo-American Telescope (AAT; \citealp{Lidman20}) using the AAOmega spectrograph. Details of the spectroscopic follow-up program are provided in \citet{Smith&DAndrea20}.

To evaluate cosmological biases, we start with the cosmological sample from \citet{Vincenzi24} and restrict the sample for this analysis to SNe whose host galaxies: i) have available spectroscopic redshifts (as a `truth' point of comparison) and ii) have available photometric redshifts (described in Section \ref{sec:host-photoz}).

\subsection{Low-$z$ SN Samples}
To measure cosmological parameters, we anchor the Hubble diagram with external low-$z$ SN samples. We use \LOWZSNe{} SNe from several smaller samples (CSP; \citealp{CSP}, CfA3; \citealp{Hicken09a}, CfA4; \citealp{Hicken12}) and \FoundationSNe{} SNe from the Foundation sample \citep{Foley18}, totaling \LOWZFOUND{}. For each of these low-$z$ samples, we use the \specz\ rather than a \photoz, as spectroscopic redshifts are expected to be readily available for such samples even for next generation surveys. 

\begin{figure*}
	\centering
	\includegraphics[width=\textwidth]{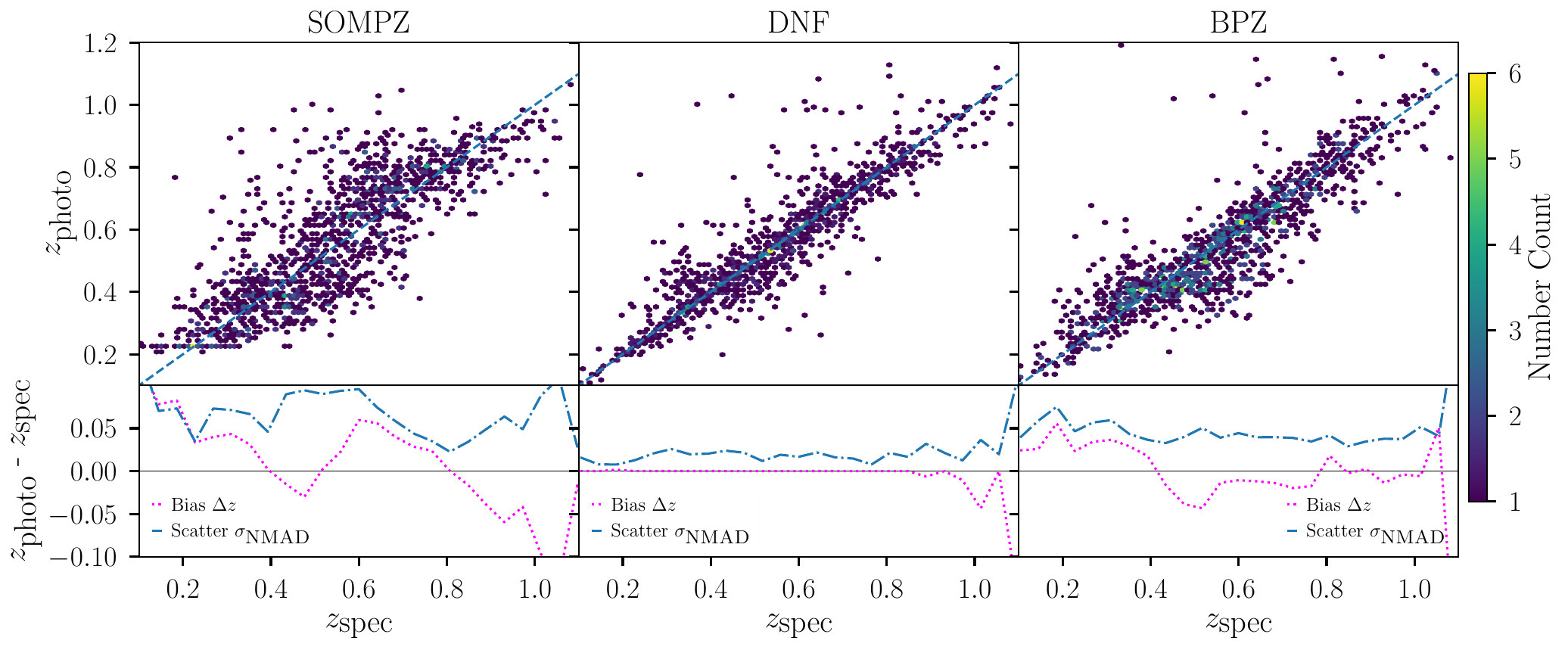}
	\caption{Top panels: host-galaxy photo-$z$ vs. spectroscopic redshifts for the $\sim14,000$ SN candidate host galaxies identified in Y3GOLD. Bottom panels: binned redshift bias ($\overline{\Delta z}$) in dotted pink and redshift scatter in dashed-dotted blue for the three \photoz\ estimates used as host-galaxy priors for the \photoz\ in this analysis: SOMPZ (left), DNF (middle), BPZ (right).}
	\label{fig:hostzbias}
\end{figure*}

\subsection{Host Galaxy Photometric Redshifts}\label{sec:host-photoz}
To estimate host galaxy \photoz, we use $griz$ photometry from the wide-field program rather than the deeper SN-field photometry. Additional near infrared (NIR) photometry ($JHKs$) has been used in a small subset of fields for \photoz\ training \citep{DESDeepFields}, but we do not use the NIR photometry in our analysis. While SN contamination can potentially bias the photometry, the $\sim$ month long transient duration is small compared to the 30 months of co-added images. 

We use host galaxy \photoz\ estimates from three different methods to capture some of the diversity of \photoz\ algorithms, as each approach is sensitive to different systematics. We focus on i) Self Organizing Map $p(z)$ (SOMPZ; \citealt{Buchs19}, \citealt{Myles21}), ii) Bayesian \photoz\ (BPZ; \citealt{BPZ}), and iii) Directional Neighbourhood Fitting (DNF; \citealt{DNF}). For SOMPZ, we consider both point estimates with Gaussian uncertainty, as well as the full PDF ($p(z)$). For BPZ and DNF, we consider only the point estimates with Gaussian uncertainty.

Because of efforts from wide field static science probes, these host-galaxy photometric redshifts are available from the Y3GOLD `value-added' catalog \citep{Y3Gold} and for the simulated galaxy catalogs we use to build our host-galaxy library (see Section \ref{sec:hostlib}). 

To evaluate the performance of photometric redshift point estimates, we consider the following standard metrics:
\begin{itemize}
    \item \photoz\ bias: $\Delta z \equiv z_{\textrm{photo}} - z_{\textrm{spec}}$
    \item \photoz\ scatter (Normalized Mean Absolute Deviation):
    \begin{equation}
        \hspace*{45pt} \sigma_{\rm NMAD} \equiv 1.48 \times \textrm{Median}\left (\frac{\left| \Delta z - \overline{\Delta z} \right|} {1+z_{\textrm{spec}}}\right) 
    \end{equation}

    \item fraction of outliers $\eta$ with $\left|\Delta z\right| > 0.1$

\end{itemize}

\subsubsection{Self-Organizing Map $p(z)$ (SOMPZ)}
SOMPZ is one of three independent \photoz\ methods used as part of the overall DESY3 3x2 pt analysis redshift scheme, along with clustering redshifts \citep{Gatti22} and shear ratios \citep{Sanchez22}. The SOMPZ method leverages information from subsamples of galaxies with deep 8-band photometry (from the DES deep fields; \citealp{DESDeepFields}) and secure redshifts (from public and private spectroscopic and many-band photometric catalogs) to determine the $n(z)$ of a broader source galaxy sample with only wide field photometry in \textit{(g)riz}. A self-organizing map is constructed for both the wide field photometry and deep field photometry; i.e.\ each galaxy is assigned a phenotype based on its colors. A synthetic source injection software (BALROG; \citealp{BALROG}) is used to inject simulated deep field galaxies into real wide field images. These wide-deep pairings are then used to weight the secure redshift information and constrain the $n(z)$ to 0.01 on the mean redshift in four tomographic bins. Further details of the method are provided in \citet{Myles21}. This method is designed specifically to precisely calibrate the overall redshift distribution of a galaxy sample to weak lensing cosmology requirements, rather than to constrain an individual galaxy point estimate or $p(z)$ PDF. As an example, the DESY3 SOMPZ analysis was restricted to \textit{riz} fluxes due to constraints on using $g$-band data specific to the DES lensing analysis. 

For each individual galaxy, we assign the weighted wide field $p(z)$ for the assigned phenotype, i.e. galaxies in the same SOM cell will have the same $p(z)$. We use the SOMPZ estimates for the DESY3 weak lensing source galaxy catalog. Both the SOMPZ $p(z)$ and point estimates are considered in our analysis, but performance metrics are given only for point estimates. In the left panel of Figure \ref{fig:hostzbias} we show the redshift bias and scatter for the SOMPZ point estimates on the subset of Y3GOLD galaxies also detected as SN candidate hosts. The median \photoz\ bias across redshift bins is \SOMPZbias{} and the median \photoz\ scatter is \SOMPZscatter{}. The outlier fraction is \SOMPZoutlier{}.

\subsubsection{Directional Neighborhood Fitting \photoz\ (DNF)}
To estimate photometric redshifts, the DNF machine-learning algorithm \citep{DNF} combines a neighbourhood fitting (NF) estimator with a modified neighborhood metric (Directional Neighborhood). In brief, the NF approach allows the algorithm to not only identify galaxies with similar observables (i.e. magnitudes) as neighbors, but also galaxies with similar \textit{relative} observables (i.e. colors). The Directional Neighborhood is defined as the product of the Euclidean and Angular Neighborhoods, i.e. the Euclidean distance and angle between vectors in multi-magnitude space. DNF combines these concepts and constructs a best-fit hyperplane to the Directional Neighborhood of a given galaxy to produce a \photoz\ $p(z)$. The redshift bias and scatter are shown in the center panel of Figure \ref{fig:hostzbias} and the median bias and scatter across redshift are \DNFbias{} and \DNFscatter{} respectively. The outlier fraction is \DNFoutlier{}. By these established metrics, DNF outperforms both SOMPZ and BPZ, although works such as \citet{Schmidt20} emphasize that \photoz\ metrics should be specific to a given science case. We note that the training sample used for the DES DNF estimates include OzDES \specz, which were primarily obtained for SN candidate host galaxies. The negligible bias reported here is therefore likely attributable to the overlap between our SN host galaxies and the DNF training sample, rather than a direct measure of the algorithm's performance.

\subsubsection{Bayesian Photometric Redshifts (BPZ)}
BPZ is a template-fitting \photoz\ estimator that produces point estimates and Bayesian posteriors \citep{BPZ}. The redshift posterior is obtained by calculating the $\chi^2$ likelihood of a set of galaxy photometry given a galaxy template at a given redshift and marginalizing over six base galaxy model templates. A magnitude-dependent luminosity prior is applied for each galaxy based on the redshift-evolving luminosity functions for elliptical, spiral, and star-burst galaxies. The estimates used for this analysis are taken from the DES Y3GOLD catalog, and further details on the DES-specific code and calibration are found in \citealp{Hoyle18}. The redshift bias and scatter are shown in the right panel of Figure \ref{fig:hostzbias} and the median bias and scatter across redshift are \BPZbias{} and \BPZscatter{} respectively. The outlier fraction is \BPZoutlier{}.

\section{Simulations}
To validate our analysis and to calculate bias corrections for known selection effects in our sample, we use catalog-level simulations generated with the SuperNova ANAlysis Software (SNANA; \citealp{Kessler09}) and orchestrated with the Pippin \citep{Pippin} software. As we include a full treatment of core-collapse contamination in the analysis, the simulations are also used to train the photometric classifier, as well as to model the core-collapse likelihood (see Section 4.3).

\subsection{Simulation Overview}
\label{sec:sim_overview}
We generate a modified version of the simulations built for the main DES-SN5YR analysis \citep{Kessler19, Vincenzi21a, DES5YR}. An overview of the simulation process is as follows: first, a time-varying source SED is generated based on the SALT3 \citep{Guy07, Guy10, Kenworthy21} spectro-photometric light-curve model, and cosmological and astrophysical effects are applied (redshift, cosmological dimming, lensing, extinction, peculiar velocities). The top-of-atmosphere SEDs are integrated across survey filters to obtain the observed fluxes. Second, measurement noise is added according to the survey observing conditions (PSF, sky noise, zeropoints). Lastly, the detection trigger is applied based on survey characteristics such as detection efficiency vs. SNR.  

To model the parent populations (SN stretch and color) as a function of host-galaxy stellar mass, we use the parameters from the main DES-SN5YR analysis \citep{Vincenzi24}, which are obtained following the method described in \citet{Popovic21}. To model the intrinsic scatter $\sigma_{\rm int}$, i.e. the post-standardization scatter in Hubble residuals, we use a dust-based model \citep{BS20} with parameters constrained with the Dust2Dust software \citep{Dust2Dust}. This results in fitted distributions of intrinsic color $c_{\rm int}$, intrinsic color-luminosity relation $\beta_{\rm int}$, and extinction model parameters ($R_V$ and $E(B-V)$) for high and low mass galaxies. The modeling is done separately for both the low-$z$ and DES samples.

For non-Ia SN populations, we use templates, rates, and luminosity functions for simulations as detailed in \citet{Vincenzi21a, Vincenzi23}. These include peculiar SN Ia populations (91bg-like, Iax; \citealp{Kessler19} and references therein) and core-collapse SNe (stripped-envelope, hydrogen-rich; \citealp{Vincenzi19}). 

SN host galaxies are simulated and associated using a galaxy catalog (or host-galaxy library; HOSTLIB) which contains at minimum a galaxy redshift and is detailed in the following section. 

\subsection{Host Galaxy Library}
\label{sec:hostlib}
The nominal DES-SN5YR analysis uses a HOSTLIB built with data \citep{Qu23a} which contains all the galaxies detected using deep coadds \citep{Wiseman20}. In this work we instead use a simulation-based HOSTLIB which allows us to consider multiple \photoz\ algorithms without having to rerun \photoz\ estimation codes. To accurately capture the multi-dimensional relationship between galaxy colors, magnitudes, and \photoz\ performance, we rely on the Buzzard suite of N-body simulations \citep{Buzzard, DeRose22} to generate our host-galaxy library. This set of N-body simulations was designed to mock the DESY3 source and lens galaxy samples and SOMPZ, DNF, and BPZ have previously been run on the resulting catalog photometry. As a result, we do not need to analytically model the \photoz\ performance ourselves. We use the v0.25 SOMPZ run on Buzzard v2.0.0 and use one million random galaxies and their properties from the catalog for the simulation HOSTLIB. We do not include host galaxy flux contribution to the SN Poisson noise. 


To model SN dependencies on host galaxy properties, we focus on the host galaxy stellar mass. We use the Buzzard host galaxy photometry and follow the Spectral Energy Distribution fitting method from \citet{Sullivan10}, following templates and assumptions made as in \citet{DES5YR}, to fit masses for the simulation HOSTLIB. We also associate SNe with their host galaxies as a function of host galaxy stellar mass based on the weighting map from \citet{Vincenzi21a} and \citet{Wiseman21}.

\subsection{Redshift Efficiency}
\label{sec:zeff}

Because we require both a host \specz\ to evaluate biases as well as a host \photoz, we pay careful attention to model the efficiency of obtaining both redshifts. Normally the redshift efficiency would be more lax (i.e. be efficient to fainter magnitude) for \photoz\ than for \specz. However, because we require a \photoz\ from the wide-field Y3GOLD catalog, which is shallower than the DES-SN survey, our \photoz\ efficiency is actually stricter (i.e. drops off at brighter magnitude) than the \specz\ efficiency. While the \specz\ efficiency is typically modeled with a ``host efficiency'' map \citep{Vincenzi21a}, here the \photoz\ efficiency is determined by the magnitude distribution of simulated Buzzard galaxies combined with the depth of the Y3GOLD catalog. In other words, the \photoz\ efficiency is built into the simulations by the availability of \photoz\ for galaxies in Buzzard and we do not explicitly model a \specz\ efficiency. 

\begin{figure}
	\centering
	\includegraphics[scale=.41]{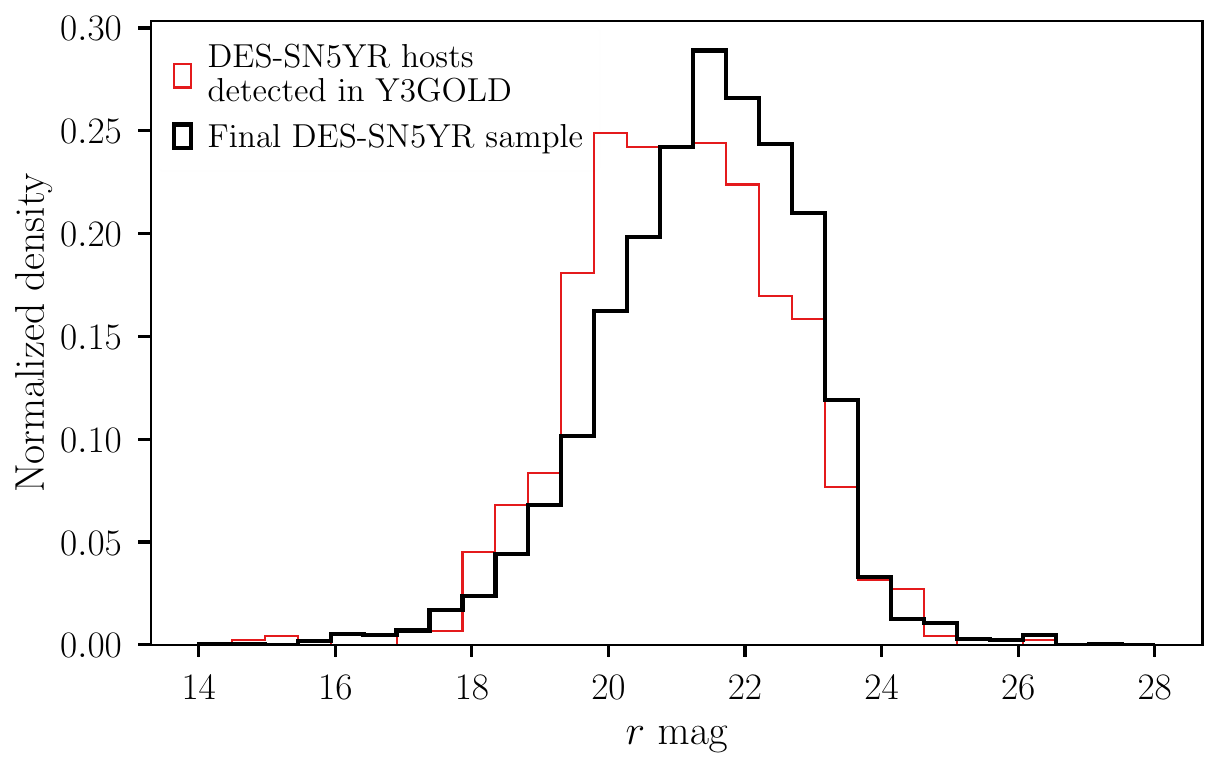}
	\caption{\textit{r}-band magnitude distributions for host galaxies of the final DES-SN5YR SN sample (black; \citealp{DES5YR}) and the DES-SN5YR host galaxies which are detected in the Y3GOLD catalog (red).}
	\label{fig:hosteff}
\end{figure}

In Figure \ref{fig:hosteff} we show the $r$-band magnitude distribution for the host galaxies in the DES-SN5YR sample (black histogram) compared to the magnitude distribution for DES-SN5YR hosts which are also detected in Y3GOLD (red histogram). This illustrates how the sample used for this analysis is skewed brighter than the nominal DES-SN5YR SN sample due to the requirement of wide field detection for \photoz.

\section{Analysis}
\label{sec:analysis}
Here we detail the analysis pipeline from light-curve fitting (Section \ref{sec:lcfit}) to final cosmology constraints (Section \ref{sec:cosmo_parameters}). 

\begin{figure*}
	\centering
	\includegraphics[width=\textwidth]{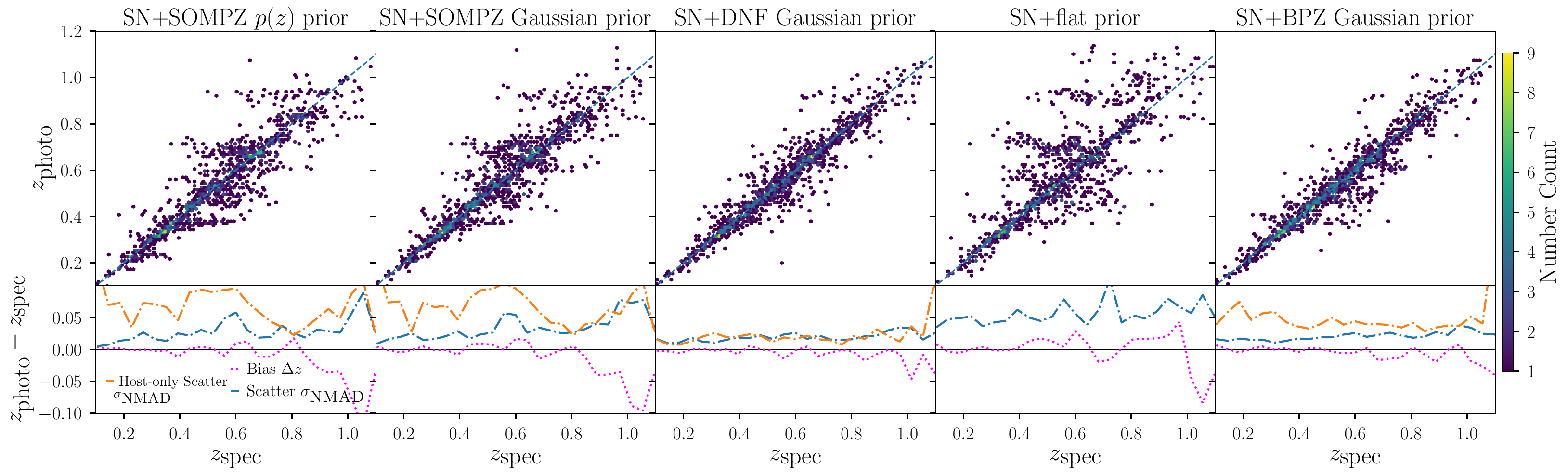}
	\caption{Top panels: SN+host photo-$z$ estimates vs. true spectroscopic redshifts. A one-to-one relation is given in blue dashed line. Bottom panels: the dotted pink lines show the binned redshift bias ($\overline{\Delta z}$) and the dashed-dotted blue lines show the redshift scatter (defined in Sec. \ref{sec:host-photoz}) across redshift bins for each SN+host \photoz\ variant. Figure \ref{fig:hostzbias} shows the equivalent for host only \photoz. The host only redshift scatter from Figure \ref{fig:hostzbias} is plotted in dashed-dotted orange lines for comparison.}
	\label{fig:datazbias}
\end{figure*}

\subsection{Light-curve Fitting}
\label{sec:lcfit}
To measure SN distances, we use the SALT3 light-curve model framework \citep{Kenworthy21}, based on SALT2 \citep{Guy10}, to fit for standardization parameters. The model is parametrized by the following parameters: $z$ (redshift), $x_{0}$ or $m_{x}$ (overall amplitude, with $m_{x} = -2.5\log_{10}(x_{0})$), $t_{0}$ (time of peak brightness), $x_{1}$ (stretch), and $c$ (color). To obtain the distance modulus $\mu$, we use the Tripp estimator \citep{Tripp}:

\begin{equation}
    \mu= m_x + \alpha x_1 - \beta c - M -\delta \mu_{\textrm{bias}} + \delta \mu_{\textrm{host}}
    \label{eq:Tripp}
\end{equation}
where $M$ is the absolute magnitude of a SN Ia with $c=0$, $x_1=0$, $\alpha$, $\beta$ are coefficients parametrizing the stretch-luminosity and color-luminosity relations, and $\delta \mu_{\textrm{bias}}$ is the bias correction applied to distances (see Section \ref{sec:BBC}). The last term, $\delta \mu_{\textrm{host}}$, is a correction for additional host-galaxy property and SN Ia property dependencies, where here we use host-galaxy stellar mass $M_*$. This correction is defined as a step function:
\begin{equation}
    \delta \mu_{\rm host} = \begin{cases}
+\gamma/2 & \quad \text{if } M_* > 10^{10} M_{\odot}, \\
-\gamma/2  & \quad \text{otherwise,}
\end{cases}
    \label{eq:massstep}
\end{equation}
where $\gamma$ is the size of the `mass step,' and $10^{10} M_{\odot}$ is the location of the `mass step' \citep{Kelly10, Lampeitl10, Sullivan10}.
For light-curve fitting, we use the implementation in SNANA, which uses a $\chi^{2}$ minimization to obtain best-fit parameters and uncertainties.

\subsection{SN+host Photometric Redshifts}
\label{sec:sn+hostz}

For this analysis, we do not use the host galaxy \photoz\ directly in place of the \specz\ but rather to help inform a \photoz\ fitted from the SN light-curve. The SALT light-curve model framework can be extended to include a fitted photometric redshift from the SN photometry; i.e. by floating the redshift rather than fixing it in the fit \citep{Kessler10a}. A host galaxy redshift prior can also be provided to improve the redshift estimates. Previous studies \citep{Mitra23} have included Gaussian host galaxy redshift priors \citep{Graham18}, with the median or average of the PDF as the mean of the Gaussian, and the RMS as the width of the Gaussian. Here, we also implement in the SNANA framework the ability to use a full host galaxy \photoz\ PDF as a prior, which should provide a more principled estimate of uncertainties and better encapsulate the degeneracies endemic to \photoz\ estimation. These PDFs are read and stored in the format of 11 quantiles in 10\% probability bins, to optimally preserve information and minimize storage space \citep{Malz18}. We consider the $p(z)$ only for SOMPZ redshift estimates, as they were previously saved in the Y3 processing and did not need to be recomputed. We include a \photoz\ variant where the SOMPZ $p(z)$ is approximated by a Gaussian to study the impact on the final SN+host \photoz\ estimate. For BPZ and DNF, we consider only point estimates with Gaussian uncertainties and refer to the final \photoz\ estimates as SN+BPZ (Gaussian) prior and SN+DNF (Gaussian) prior.

We note a subtlety related to the implementation of this \photoz\ fitting in SNANA; if the SED model range does not extend sufficiently into bluer wavelengths, this may cause filter dropouts at high redshift. This can result in artificially small $\chi^2$ values that cause low redshift SNe to be pathologically fit with high redshift values. To avoid this problem, we use the SALT3 extended wavelength model with fit range 2000-13000 $\AA$ and linear extrapolation to zero flux at $\lambda = 500\AA$. 

In Figure \ref{fig:datazbias} we show the SN+host \photoz\ estimates plotted against true spectroscopic redshifts. We show the redshift bias and scatter for each SN+host \photoz\ variant in magenta and blue in the bottom panels. We also show the binned redshift scatter for each host-galaxy \photoz\ prior (without SN information) in orange in the bottom panels. Except for the case of SN+DNF prior, the redshift scatter obtained from host galaxy photometry only is reduced by $\sim50\%$ when adding SN information. This indicates that adding SN data significantly improves the \photoz\ estimate. In the SN+DNF case, the redshift scatter is not improved from the DNF host galaxy \photoz\ because the host \photoz\ information is much more precise compared to the SN. 

The best performing variant by the \photoz\ bias and scatter metrics is SN+DNF prior with the median redshift bias and scatter across redshift bins being \SNDNFbias{} and \SNDNFscatter{} respectively. We also find that the final SN+host \photoz\ is not significantly improved by using the full SOMPZ $p(z)$ prior compared to a Gaussian approximation. We discuss other potential uses for the $p(z)$ information in Section \ref{sec:discussion}. In particular, we note that the SN+SOMPZ $p(z)$ prior, SN+SOMPZ Gaussian prior, and SN $z$ fit with flat prior show a redshift-dependent bias at $z \gtrsim 0.8$, which may be caused by asymmetric migration due to the redshift cut at $z=1.2$ such that events with $z_{\textrm{photo}} < z_{\textrm{spec}}$ remain in the sample but events with $z_{\textrm{photo}} > z_{\textrm{spec}}$ are rejected.

\subsection{Sample Selection}
To select our sample, we apply standard cosmological quality cuts as follows:
\begin{itemize}
    \item fitted color $\left| c \right| < 0.3$
    \item fitted stretch $\left| x_1 \right| < 3.0$
    \item fitted stretch uncertainty $\sigma_{x_1} < 1.0$
    \item fitted $t_0$ uncertainty $\sigma_{t_0} < 2.0$ days
\end{itemize}

\renewcommand{\arraystretch}{1.5}

In Table \ref{tab:cuts} we show the number of SNe remaining after sequentially applying each cut and for different host-galaxy \photoz\ priors. In brackets, we show the percentage of SNe lost from each cut. We note that the \specz\ sample is not exactly identical to the main DES-SN5YR sample, due to several differences including i) requiring host \photoz, ii) requiring \photoz\ fit convergence and iii) using an extended SALT3 model for light-curve fitting (see Section \ref{sec:sn+hostz}) to be consistent with the \photoz\ cases.

\begin{table*}
  \centering
\begin{tabular}{l|l|p{1.8cm}|p{1.8cm}|p{1.8cm}|p{1.8cm}|p{1.8cm}|}
& \multicolumn{6}{c}{\textbf{$\#$ SNe for Redshift Method [$\%$ cut]}} \\
\cline{2-7}

\textbf{Cut} & \specz\ & SN+SOMPZ \newline $p(z)$ prior & SN+SOMPZ \newline Gaussian prior & SN+DNF \newline Gaussian prior & SN+flat prior & SN+BPZ \newline Gaussian prior\\
\hline \hline
Host-galaxy \photoz\ available & 13,793 & 13,793 & 13,793 & 13,793 & 13,793 & 13,793\\
\hline
SALT3 fit converged                         & 2377 &         2427 &         2392 &         2392 &         2400 &         2392 \\
\hline
SN color $\left| c \right| < 0.3$           & 2087 [13.9\%] & 2131 [13.9\%] & 2114 [13.2\%] & 2114 [13.2\%] & 2132 [12.6\%] & 2114 [13.2\%] \\
\hline
SN stretch $\left| x_1 \right| < 3.0$       &  1902 [9.7\%] & 1817 [17.3\%] & 1857 [13.8\%] & 1857 [13.8\%] & 1777 [20.0\%] & 1857 [13.8\%] \\
\hline
SN stretch uncertainty $\sigma_{x_1} < 1.0$ & 1623 [17.2\%] & 1547 [17.5\%] & 1562 [18.9\%] & 1562 [18.9\%] & 1491 [19.2\%] & 1562 [18.9\%] \\
\hline
SN $t_0$ uncertainty $\sigma_{t_0} < 2.0$   &  1614 [0.6\%] &  1537 [0.7\%] &  1552 [0.6\%] &  1552 [0.6\%] &  1484 [0.5\%] &  1552 [0.6\%] \\
\hline
Valid bias correction                       &  1577 [2.3\%] &  1481 [3.8\%] &  1505 [3.1\%] &  1506 [3.1\%] &  1420 [4.5\%] &  1474 [5.3\%] \\
\hline \hline
\textbf{Final sample size}                           &         \textbf{1577} &         \textbf{1481} &         \textbf{1505} &         \textbf{1506} &         \textbf{1420} &         \textbf{1474} \\
\end{tabular}
\caption{Summary Table of SN sample cuts (DES-only)}
\label{tab:cuts}
\end{table*}

\begin{figure}
	\centering
	\includegraphics[width=\columnwidth]{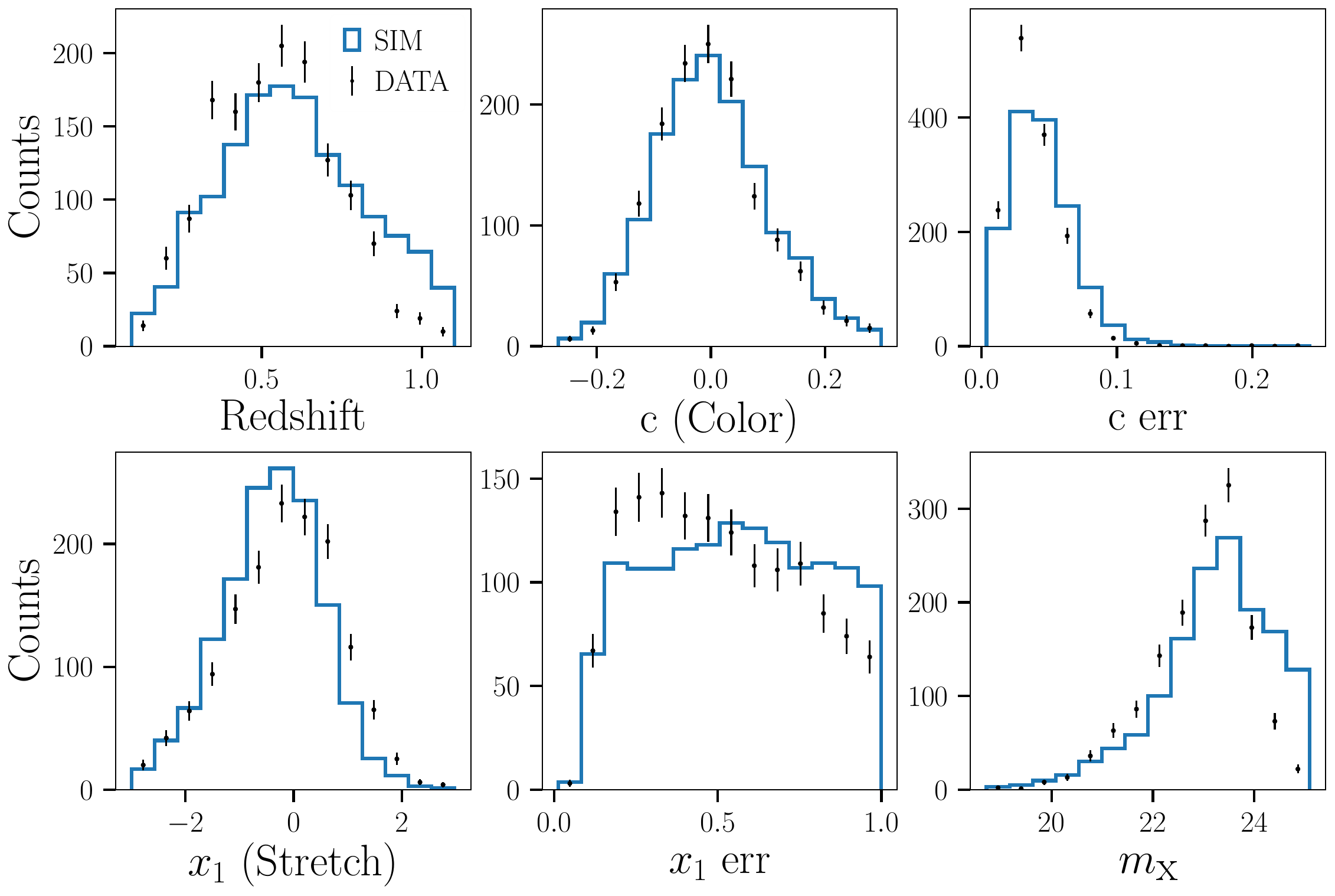}
	\caption{Distributions of \specz, fitted SN light-curve parameters using \specz\ ($c$, $x_1$, $m_B$), and their errors for data (black points) and simulations (blue histogram). The simulation histograms are normalized to the number of data points.}
	\label{fig:sim_data}
\end{figure}

\begin{figure}
	\centering
	\includegraphics[width=\columnwidth]{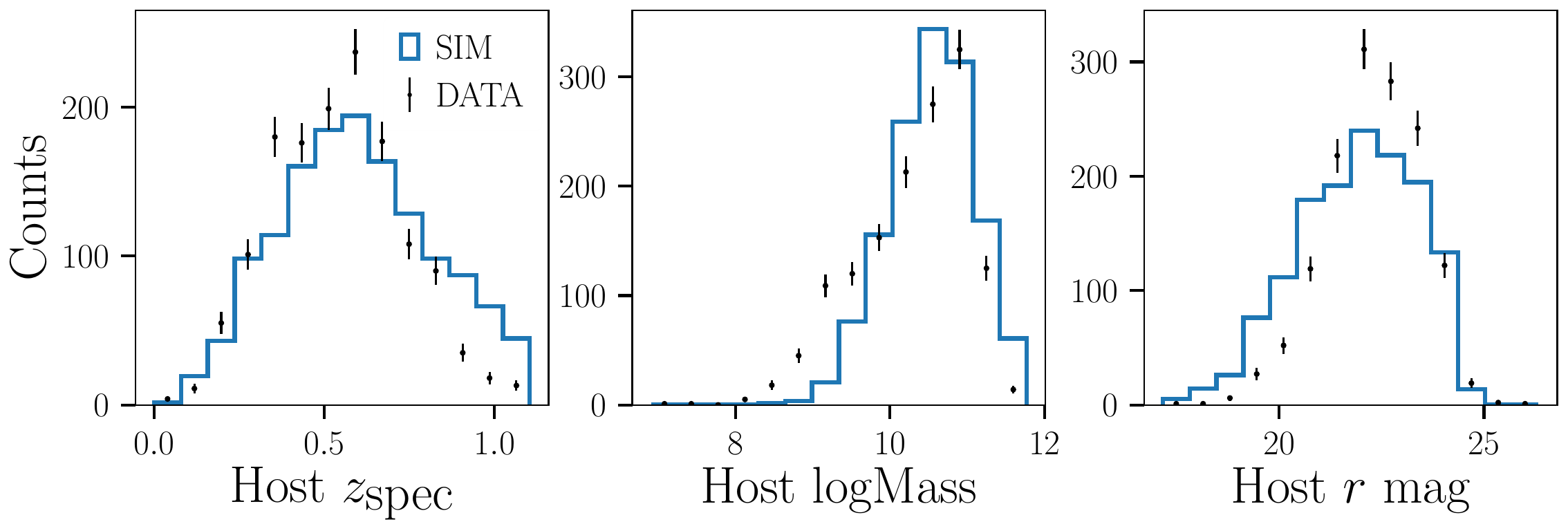}
	\caption{Distributions of SN host-galaxy properties for data (points) and simulations (blue histogram).}
	\label{fig:sim_data_host}
\end{figure}

\subsection{Data and Simulation Comparison}
\label{sec:datasimcomp}

In Figure \ref{fig:sim_data} we show the distributions comparing simulations and data for fitted SALT3 light-curve parameters and their uncertainties, normalized to the number of SNe in the data. We also compare the distributions for simulations and data for host-galaxy properties in Figure \ref{fig:sim_data_host}. The agreement in host magnitude in particular illustrates that the ``\photoz\ selection function'' is sufficiently reproduced, although the fainter tails for the simulation redshift and $m_x$ distributions also indicate that this modeling can be improved. We emphasize that correctly modeling and propagating the relationship between galaxy properties and redshift biases is a crucial component of the analysis. In Appendix A, we show the equivalent light-curve parameter comparison plots for when the redshift is floated simultaneously with varying host-galaxy priors. We manually scale down the simulated total number of SNe, as well as the rate of core-collapse contamination, to better match the data, as we are unable to replicate the same values from first principles alone.

\subsection{SN Classification}
\label{sec:classification}
To classify our supernova sample, we use the photometric classifier SuperNNova (SNN, \citealt{Moller20}), a recurrent neural network trained on simulated light-curves. Previous analyses have rigorously evaluated the performance of SNN for the DES-SN5YR analysis, with a prediction of less than 1.5 percent contamination in the final sample \citep{Vincenzi21a, Moller22}. We train a model of SNN on our nominal Ia and non-Ia simulated light-curves as detailed in Section \ref{sec:sim_overview}.

\subsection{BEAMS with Bias Corrections}
\label{sec:BBC}

We use an extension of the Bayesian Estimation Applied to Multiple Species (BEAMS; \citealp{Kunz07}), BEAMS with Bias Corrections (BBC; \citealp{BBC}) to address core-collapse contaminants and to correct for known selection effects and Malmquist bias for our SN sample. The BEAMS method incorporates probabilistic information from a photometric classifier to marginalize over core-collapse contamination. In the BEAMS framework, the likelihood is modeled with two terms, one for the SN Ia population (modeled analytically), and one for core-collapse contaminants (empirically modeled using simulations; see Section \ref{sec:sim_overview}). Using SNN, each SN in the sample is assigned a probability of being a Ia, with zero being non-Ia and one being likely-Ia. These probabilities are then used to weight contributions to the Hubble diagram. 

BBC extends the BEAMS framework and incorporates the method from \citet{Marriner11} to determine standardization nuisance parameters in a cosmology-independent way. The BBC fit determines global nuisance parameters ($\alpha$, $\beta$, $\gamma$, $\sigma_{\rm int}$) and also determines bias-corrected distances in redshift bins to create a binned Hubble diagram. When including systematic uncertainties, an unbinned Hubble diagram approach results in smaller uncertainties than a binned approach \citep{PantheonPlus,Kessler23}, but for a statistical uncertainty only approach, there is no difference. As we do not include a full systematic covariance matrix in this analysis, we use the binned BBC approach rather than unbinned or re-binned. 

For bias corrections, BBC applies a distance correction based on the true and measured distances for a large set of survey-specific simulations ($\delta \mu_{\rm bias}$ in Equation \ref{eq:Tripp}). We use the BBC-4D framework \citep{Popovic21}, which computes the bias corrections in bins of \{$z, x_1, c, \log{M_*}$\} to be compatible with dust-based models. In our analysis, we use the respective measured redshift in each case (e.g. \specz, SN+SOMPZ Gaussian prior, SN+DNF Gaussian prior, etc.) to compute the corrections. The bias corrections for the low-$z$ sample are computed separately following \citet{Scolnic18}, \citet{Kessler19}, and \citet{DES5YR}. We also compute the bias correction $\sigma_{\rm int}$ separately by sample to prevent an overinflated $\sigma_{\rm int}$ for the low-$z$ and Foundation samples, which use \specz.

\subsection{Distance Uncertainties}
\label{sec:mu_uncert}
The individual distance modulus uncertainty for a SN in BBC (with the $i$ index ignored for clarity) is given by \citet{BBC, PantheonPlus}:

\begin{equation}
 \begin{aligned} \label{eq:mu_uncertainty}
    \sigma_{\mu}^2 = &f(z, c, M_*)\sigma_{\rm SALT3}^2 + (\sigma_{\mu}^{\textrm{vpec}})^2 + (\sigma_{\mu}^z)^2 + \sigma_{\rm lens}^2 \\
 \end{aligned}
\end{equation}

The light-curve fit parameter uncertainty $\sigma_{\rm SALT3}^2$ is given by:

\begin{equation}
 \begin{aligned}
    \sigma_{\rm SALT3}^2 = C_{m_B,m_B} + \alpha^2C_{x1,x1} + \beta^2C_{c,c}  + 2\alpha C_{m_B,x1} \\ -2\beta C_{m_B,c} - 2\alpha\beta C_{x1,c},\\
 \end{aligned}
\end{equation}
where $C$ is the fitted covariance matrix for the light-curve parameters. $\sigma_{\mu}^{\textrm{vpec}}$ is the contribution from the peculiar velocity uncertainty $\sigma_{\textrm{vpec}}$ given by:

\begin{equation}
  \sigma_{\mu}^{\textrm{vpec}} = \left(\frac{5}{\ln(10)}\right) \frac{1+z}{z(1+z/2)} \left(\frac{\sigma_{\textrm{vpec}}}{c}\right) ,
\end{equation}
as explained in \citet{Davis11}, and $\sigma_{\mu}^z$ is the contribution from the redshift uncertainty $\sigma_z$. We set $\sigma_{\mu}^z = 0$, as $\sigma_z$ already inflates uncertainties on $m_x, x_1$, and $c$ when redshift is floated in the light-curve fit; therefore its inclusion causes an overestimated uncertainty due to the correlated color error (see \citealp{Chen22} Appendix A.1 for details). The $f(z, c, M_*)$ error scaling factor is introduced following the original BBC and determined from simulations such that rms($(\mu - \mu_{\rm true})/\sigma_{\mu}) = 1$ in \{$z, c, M_*$\} bins. The $f(z, c, M_*)$ scaling is needed because the naively computed uncertainty does not account for Malmquist bias that removes faint events and thus reduces the scatter (and uncertainty) at higher redshifts. 

We do not include an error floor term $\sigma_{\rm floor}^2 (z,c,M_*)$ that was included in the main DES-SN5YR analysis. We discuss the subtleties related to this distance uncertainty error scaling when using \photoz\ in Section \ref{sec:results}. Details concerning the justification and implementation of the error corrections can be found in \citet{BBC} and \citet{Vincenzi24}.

\subsection{Cosmological Parameter Inference}
\label{sec:cosmo_parameters}

To constrain cosmological parameters, the $\chi^2$ of the SN likelihood is given as

\begin{equation}
\label{eq:chi2}
\chi^2 = \Delta\mu^T \cdot C_{\rm stat}^{-1} \cdot \Delta\mu,
\end{equation}

where $\Delta \mu$ is the data vector of distance measurements $\{ \mu_{\textrm{obs,}i} - \mu_{\textrm{theory,}i}(\Omega_M, w)\}_{i=1,...,N_{\textrm{SNe}}} $ and $C_{\rm stat}$ is the statistical-only uncertainty covariance matrix as defined in \citet{DES5YR}.

We use the ``{\tt wfit}'' $\chi^{2}$ minimization program as implemented in SNANA with a prior from the Cosmic Microwave Background (CMB) R-shift parameter, which is derived from simulations with the same underlying cosmology as our simulated samples. We fix the R-shift parameter uncertainty to $\sigma_R = 0.006$ following the constraining power of \citet{Planck2020}.

\section{Results}
\label{sec:results}

We perform an identical analysis following the framework presented in Section \ref{sec:analysis} using spectroscopic redshifts and using photometric redshifts, where the \photoz\ variations are:
\begin{itemize}
    \item SN + SOMPZ $p(z)$ prior
    \item SN + SOMPZ Gaussian prior
    \item SN + DNF Gaussian prior
    \item SN + BPZ Gaussian prior
    \item SN + flat prior

\end{itemize}

To isolate the biases that result from using \photoz, we consider the difference in constraints on $w$ when using \photoz\ and \specz\ defined as:
\begin{equation}
    \Delta w \equiv w_{\textrm{\photoz}} - w_{\textrm{\specz}},
\end{equation}
where $w_{\textrm{\specz}}$ is the constraint obtained using spectroscopic redshifts and $w_{\textrm{\photoz}}$ is the constraint obtained using a given photometric redshift variant. 

We note that the set of SNe used to obtain $w_{\textrm{\specz}}$ and $w_{\textrm{\photoz}}$ are not identical (see Table \ref{tab:cuts}). We choose not to use a uniform SN sample across \photoz\ variations, as in the future we will not have \specz\ for all SNe and therefore will not be able to approximate a selection based on which SNe would pass cuts when using \specz. Instead, we take a realistic approach and allow each redshift case to be selected by our conventional cuts (see Table \ref{tab:cuts}). In other words, when calculating $\Delta w$ values the \specz\ sample size is fixed to the larger value while the \photoz\ sample sizes vary. 

To validate our analysis methods, we verify with simulations that when using \specz\ with identically generated bias corrections, we recover the true cosmology of $w=-1$ to within 0.01. In Section \ref{sec:popnuisance} we discuss the impact of using \photoz\ on the SN light-curve parameters and BBC nuisance parameters. In Section \ref{sec:simresults} we present results for simulations and in Section 
\ref{sec:dataresults} we present results for data. In Section \ref{sec:errorscaling} we describe subtleties related to error scaling for distance modulus uncertainties.

\begin{table*}
\centering
\begin{tabular}{l|c|c|c|c|c|c|c|c|}
& \multicolumn{4}{l}{\textbf{Data}}  & \multicolumn{4}{l}{\textbf{Simulations Ia+CC}} \\
\hline \hline
\textbf{Redshift}                & $\alpha$ & $\beta$ & $\gamma$ & $\sigma_{\rm int}$ & $\alpha$ & $\beta$ & $\gamma$ & $\sigma_{\rm int}$\\
\hline
spec-$z$ & 0.159$\pm$0.004 & 3.274$\pm$0.034 &  0.026$\pm$0.009 & 0.102 & 0.140$\pm$0.003 & 3.087$\pm$0.046 & -0.015$\pm$0.009 & 0.078\\
SN+SOMPZ $p(z)$ prior & 0.138$\pm$0.000 & 2.821$\pm$0.002 &  0.037$\pm$0.010 & 0.162 & 0.142$\pm$0.005 & 3.089$\pm$0.063 & -0.014$\pm$0.015 & 0.157\\
SN+SOMPZ Gaussian prior & 0.137$\pm$0.004 & 2.885$\pm$0.004 &  0.033$\pm$0.010 & 0.153 & 0.142$\pm$0.006 & 3.097$\pm$0.061 & -0.015$\pm$0.016 & 0.150\\
SN+DNF Gaussian prior & 0.146$\pm$0.000 & 3.027$\pm$0.000 &  0.004$\pm$0.009 & 0.131 & 0.142$\pm$0.004 & 3.085$\pm$0.051 & -0.012$\pm$0.013 & 0.118\\
SN+flat prior & 0.147$\pm$0.003 & 2.972$\pm$0.031 &  0.009$\pm$0.010 & 0.175 & 0.142$\pm$0.005 & 3.066$\pm$0.055 & -0.018$\pm$0.013 & 0.169\\
SN+BPZ Gaussian prior & 0.156$\pm$0.004 & 3.076$\pm$0.044 & -0.019$\pm$0.010 & 0.198 & 0.143$\pm$0.005 & 3.065$\pm$0.039 & -0.017$\pm$0.012 & 0.202\\
\end{tabular}
\caption{BBC nuisance parameter values for data and simulations}
\label{tab:deltanuisance}
\end{table*}

\subsection{Impact on light-curve and nuisance parameters}
\label{sec:popnuisance}
\begin{figure*}
	\centering
	\includegraphics[scale=.35]{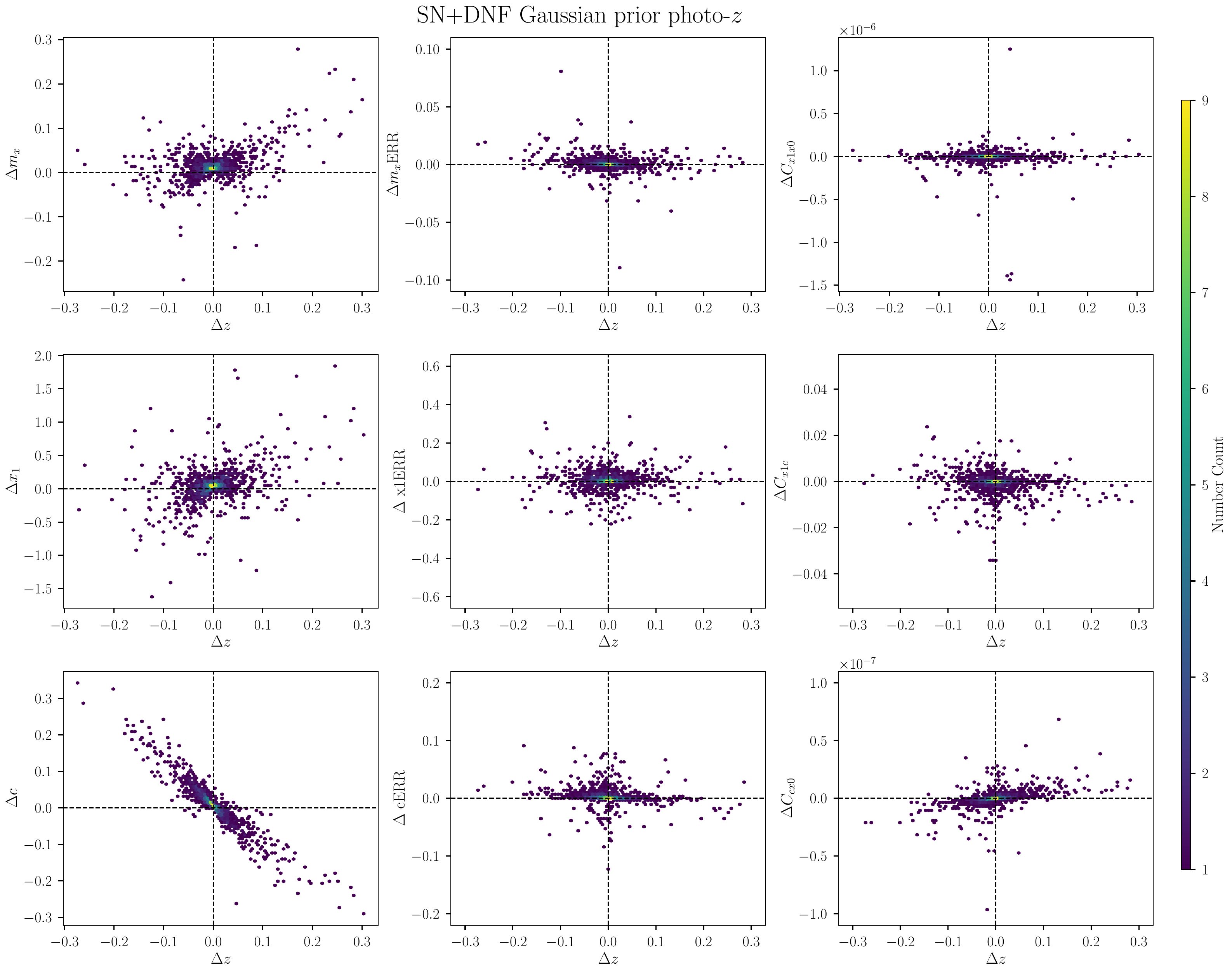}
	\caption{Bias in fitted light-curve parameters ($m_x, x_1, c$), their errors ($m_x$ERR, $x_1$ERR, $c$ERR), and their covariances ($C_{x1,x0}, C_{x1,c}, C_{c,x0}$) as a function of $z$ bias in the data using SN+DNF Gaussian prior photo-$z$.}
	\label{fig:lcbias}
\end{figure*}

In Figure \ref{fig:lcbias} we show how fitted light-curve parameters, their errors, and their covariances, are changed as a function of the \photoz\ bias for the entire data sample. In Table \ref{tab:deltanuisance}, we show the fitted BBC nuisance parameters $\alpha,\beta,\gamma$, and $\sigma_{\rm int}$ for each redshift case, for data and simulations. The errors for data are given from the BBC fit, while the errors for simulations are given as the standard deviation of the 25 simulation instances. The true input $\alpha$ and $\gamma$ are 0.145 and 0 respectively, while the input $\beta$ is not comparable with the fitted $\beta$ (see \citealp{Dust2Dust} for further explanation on the difference between $\beta_{SN}$ and $\beta_{\rm SALT}$).

Most notably, biases in redshift ($\Delta z$) are strongly correlated with biases in SN color ($\Delta c$). As a result, the color-luminosity relation $\beta$ is also biased with respect to the values obtained using \specz. As explained in \citet{Chen22}, these effects do not cause a bias in cosmology because the $\Delta \mu$ calculated from $\Lambda$CDM for small $\Delta z$ values is roughly equal to $-\beta \Delta c$ such that SNe are self-corrected toward the fiducial cosmology (see also Sec 5.3 of \citealp{Mitra23}). However, this effect is not propagated to uncertainty in nuisance parameters and may lead to a drastically underestimated error on $\beta$ as is seen for the data. Further, the recovered $\beta$ is consistent across each redshift case in the simulations but fluctuates in the data.

We recover a mass step $\gamma$ ($\sim0.03$) in the data nearly consistently across redshift cases, whereas in the simulations, we recover a mass step consistent with zero for each redshift. This is consistent with what is seen in the DES-SN5YR analysis, indicating that a dust-based intrinsic scatter model alone is not sufficient to completely explain the mass step (see also \citealp{Wiseman22, Kelsey23}). We also note that $\alpha$ is consistent between redshift cases in the simulations, but biased by $\sim 0.02$ when using \photoz\ for the data.

For both data and simulations, the intrinsic scatter is notably increased when using \photoz. While this is an indication that our method of accounting for redshift uncertainty contribution to distance uncertainties could be improved, we leave this investigation to future work. 

\begin{table*}
\centering
\begin{tabular}{l|c|c|c|c|c|c|c|c|c|c}
& \multicolumn{2}{l}{\textbf{Data}}  & \multicolumn{4}{l}{\textbf{Simulations Ia+CC}}  & \multicolumn{4}{l}{\textbf{Simulations Ia Only}}\\
\hline \hline
\textbf{Redshift}                & $\Delta w$ & $\sigma_{w}$ & $\Delta w$ & $\Delta w$ Error & $\sigma_{w}$ & $\Delta w$ RSD & $\Delta w$ & $\Delta w$ Error & $\sigma_{w}$ & $\Delta w$ RSD  \\
\hline
                 spec-$z$ &              0.000 &            0.023 &                   0.000 &                         0.000 &                 0.020 &                       0.000 &                    0.000 &                  0.020 &                          0.000 &                        0.000 \\
    SN+SOMPZ $p(z)$ prior &             -0.019 &            0.026 &                  -0.007 &                         0.003 &                 0.024 &                       0.014 &                   -0.005 &                  0.024 &                          0.003 &                        0.014 \\
SN+SOMPZ Gaussian prior &             -0.002 &            0.025 &                  -0.006 &                         0.002 &                 0.024 &                       0.012 &                   -0.009 &                  0.024 &                          0.002 &                        0.010 \\
  SN+DNF Gaussian prior &             -0.017 &            0.025 &                  -0.003 &                         0.002 &                 0.022 &                       0.009 &                   -0.002 &                  0.022 &                          0.002 &                        0.009 \\
          SN+flat prior &             -0.035 &            0.027 &                   0.004 &                         0.002 &                 0.034 &                       0.011 &                    0.013 &                  0.034 &                          0.003 &                        0.015 \\
  SN+BPZ Gaussian prior &             -0.038 &            0.029 &                   0.007 &                         0.003 &                 0.025 &                       0.013 &                    0.006 &                  0.025 &                          0.002 &                        0.012 \\
\end{tabular}
\caption{Difference in $w$ ($\Delta w$) values between \specz\ and each \photoz\ case}
\label{tab:deltaw}
\end{table*}

\begin{figure*}
	\centering
	\includegraphics[width=\textwidth]{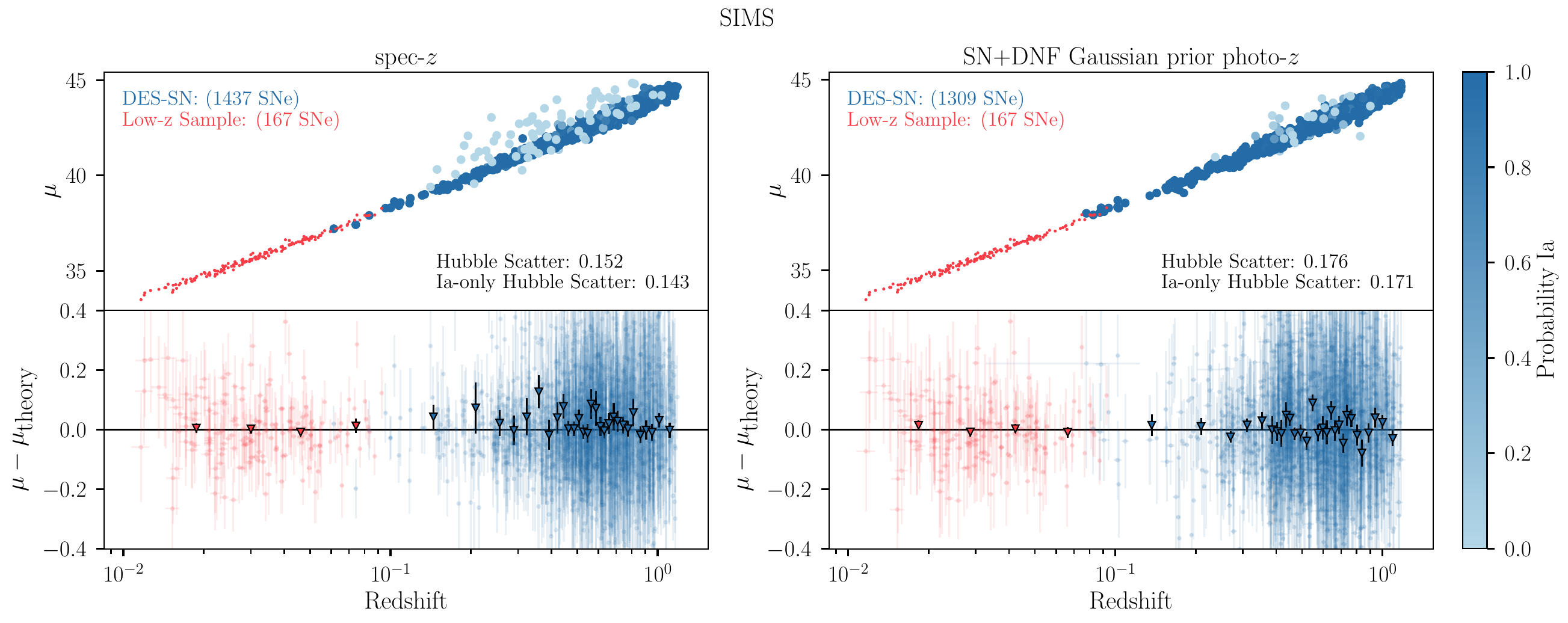}
	\caption{Hubble diagram from analyzing simulated data (upper panels) and Hubble residuals (lower panels; w.r.t a flat $\Lambda$CDM cosmology) using spec-$z$ (left) and SN+DNF prior \photoz\ (right). DES SNe are shown in blue, and low-$z$ SNe are shown in red. The Hubble residual means in redshift bins are overplotted in the lower panels as triangles. The DES-SN sample is color coded by probability of being an SN Ia as given by SNN.}
	\label{fig:simHD_DNF}
\end{figure*}

\begin{figure}
	\centering
	\includegraphics[width=\columnwidth]{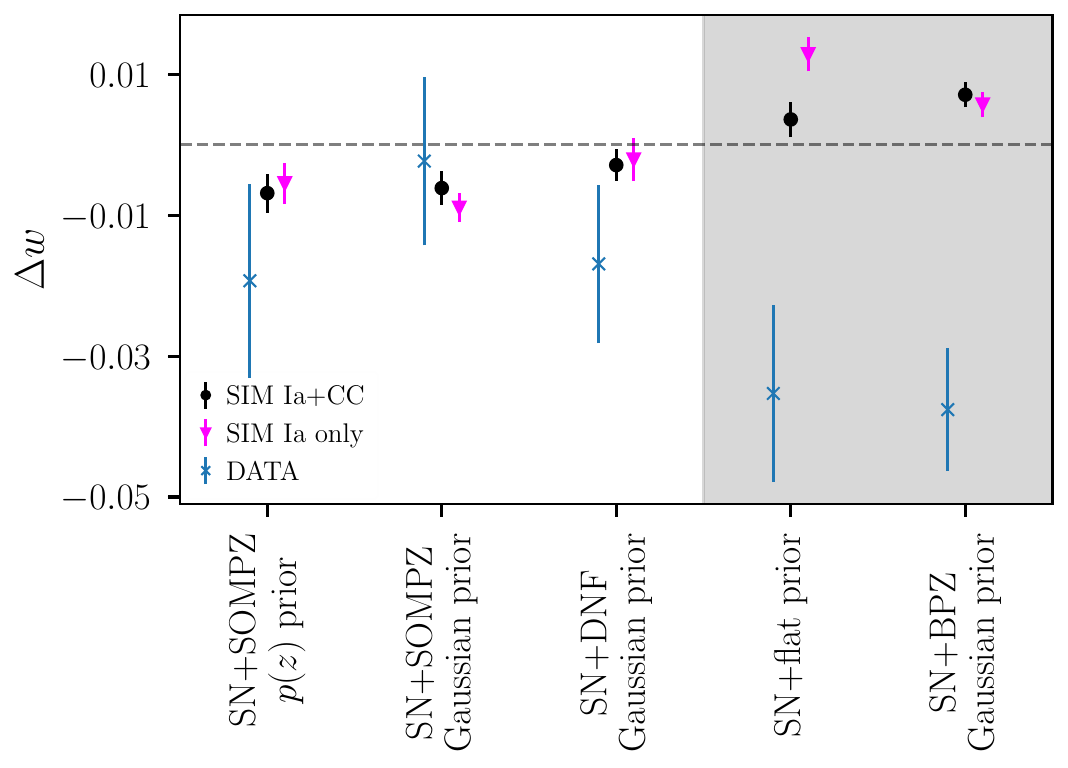}
	\caption{$\Delta w = w_{\textrm{\photoz}} - w_{\textrm{\specz}}$ values for simulations with Ia+CC SNe (black circle) and Ia Only (magenta triangle) with the standard error on $\Delta w$. $\Delta w$ values for data are shown as blue X's with standard deviation (RSD in Table \ref{tab:deltaw}) from 25 instances of simulations. The greyed area indicates \photoz\ cases which have significantly discrepant results between the simulations and data.}
	\label{fig:deltaw_ccia}
\end{figure}

\subsection{Cosmology Results from Simulations}
\label{sec:simresults}
For the \specz\ and each \photoz\ case, we generate 25 statistically independent Hubble diagrams with bias corrected distances, with $\sim$ \nsimSNe{} SNe in the simulated samples. In the top panels of Figure \ref{fig:simHD_DNF}, we show one of the 25 Hubble diagrams when using \specz\ and the SN+DNF prior \photoz, which has the lowest redshift scatter. In the bottom panel we plot the binned and unbinned Hubble residuals, the difference between our measured distances and distances from the best fit $w$CDM cosmology, as a function of redshift. The \specz\ case has Hubble scatter, i.e. the rms scatter of the Hubble residuals, of \simSPEChscatter{} while the SN+DNF prior \photoz\ has Hubble scatter \simDNFhscatter{}. The Hubble diagrams for the remaining \photoz\ cases are shown in Appendix B, along with the Hubble scatter in each panel. We find that the Hubble scatter is larger than the \specz\ case across each \photoz\ method as expected. 

We provide a summary of the resulting $\Delta w$ values (with respect to \specz) in Table \ref{tab:deltaw} for simulations including non-Ia SNe (Ia+CC, middle section) and for Ia-only simulations (right section). Each $\Delta w$ value is averaged over the 25 statistically independent realizations of the simulation. The robust standard deviation (RSD; $1.48\times \rm Median(\left| \Delta w \right|)$) and standard error (RSD/$\sqrt{25}$) of the $\Delta w$ values are also given for each case. We also report the average uncertainty on $w$, $\sigma_w$, as a point of comparison for the significance of the given $\Delta w$ values. We find $\Delta w$ values $< |0.01|$ for each \photoz\ method. In particular, SN+DNF prior, SN+SOMPZ $p(z)$ and SN+SOMPZ Gaussian prior are able to recover the true $w$ to \simDNFdw{}, \simSOMPZPZdw{}, and \simSOMPZGdw{} respectively. We further discuss in Section \ref{sec:dataresults} why SN+flat prior and SN+BPZ prior are disfavored despite their small $\Delta w$ values.

Given that the SN+SOMPZ $p(z)$ and SN+SOMPZ Gaussian prior estimates have higher redshift scatter as seen in Figure \ref{fig:datazbias}, it is somewhat unexpected that their use results in a small bias. SOMPZ estimates are also calibrated specifically to provide unbiased redshift \textit{distributions} for wide tomographic bins (and as such the $p(z)$ for individual galaxies are principled but wide). Our ability to infer unbiased cosmology using SOMPZ estimates in our current framework, which requires a point estimate redshift, is of note. We discuss further potential applications for $p(z)$ in Section \ref{sec:discussion}. We also note that using SN+SOMPZ $p(z)$ and SN+SOMPZ Gaussian prior result in significantly more lost events compared to the other \photoz\ cases.

\subsubsection{Photometric Classification}
To understand the interacting effects of photometric classification with \photoz, we consider simulations in which each SN is classified perfectly, i.e. cosmology is constrained only using SNe Ia. For simulations containing only SNe Ia, the $\Delta w$ values shift only marginally compared to simulations with both Ia+CC, on average across cases by \simCCdw{}. While the $\Delta w$ values for Ia+CC sims are consistently larger than for Ia only sims, the shifts are small and comparable to those reported in \cite{DES5YR}. In Figure \ref{fig:deltaw_ccia}, we show the $\Delta w$ values and their standard errors for the Ia+CC sims and Ia only sims for each \photoz\ variant. In each case, except for SN+flat prior, the change in $\Delta w$ with and without CC SNe are within $1 \sigma$ of each other, showing that core-collapse contamination is a minor systematic, even when combined with photometric redshifts. 

\subsection{Cosmology Results from Data}
\label{sec:dataresults}
\begin{figure*}
	\centering
	\includegraphics[width=\textwidth]{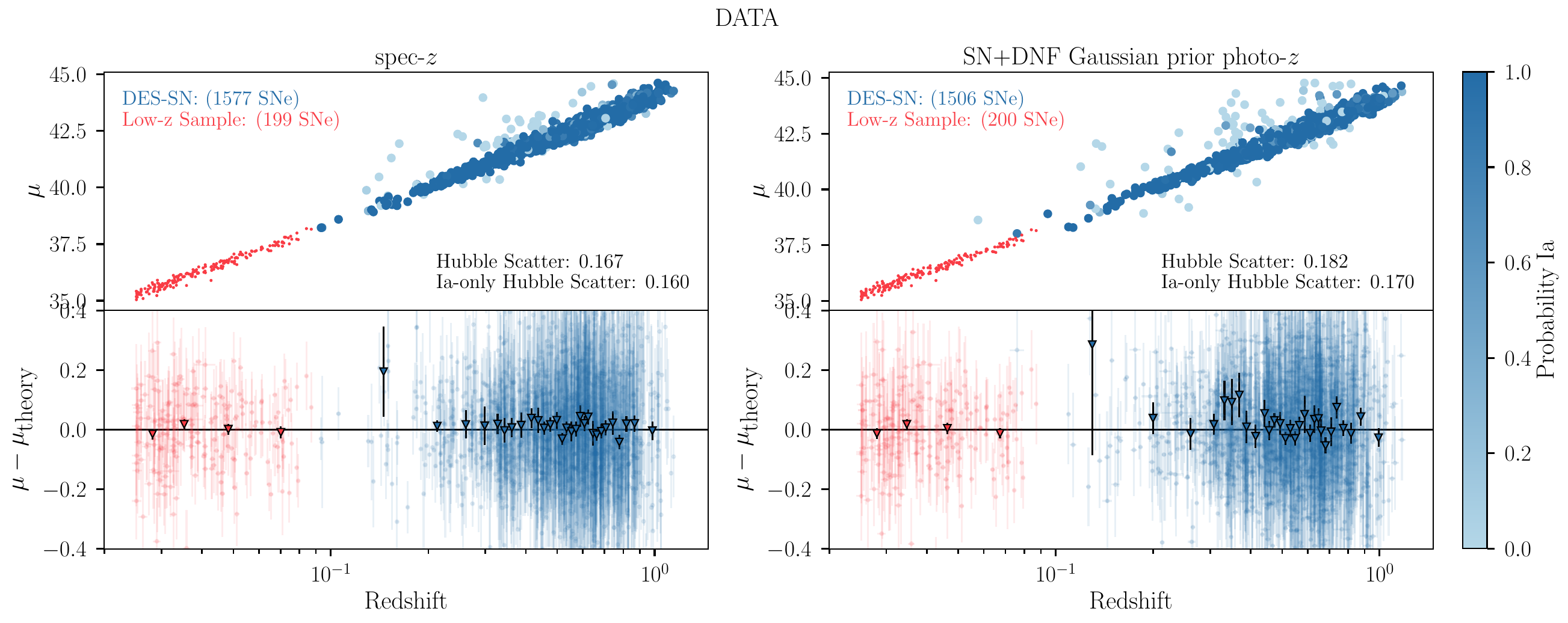}
	\caption{Same as Figure \ref{fig:simHD_DNF} but for data.}
	\label{fig:dataHD_DNF}
\end{figure*}

Using the same analysis pipeline as the simulations, we generate Hubble diagrams and fit cosmology using the dataset described in Section \ref{sec:data}. We show the Hubble diagrams for \specz\ and SN+DNF \photoz\ in the top panels of Figure \ref{fig:dataHD_DNF} and Hubble residuals as a function of redshift in the bottom panels. Hubble diagrams for the remaining \photoz\ cases are given in Appendix B. As we do not have truth values for classification in the data, the Ia-only subsample is defined as SNe having SNN probability of being a Ia $> 0.5$. 

The Hubble scatter across each \photoz\ case is larger than the \specz\ case as expected from simulations. However, the Hubble scatter for the \specz\ case is \dataSPEChscatter{}, $0.015$ larger than the sims (\simSPEChscatter{}) and $0.017$ larger for the Ia-only scatter (\dataIASPEChscatter{} vs. \simIASPEChscatter{}). This trend in Hubble scatter is not seen coherently for each \photoz\ case, where the scatter is larger in only two out of five cases in the data compared to the sims.

We find $\Delta w$ values of \dataSOMPZPZdw{} and \dataSOMPZGdw{} for SN+SOMPZ $p(z)$ prior and SN+SOMPZ Gaussian prior respectively and \dataDNFdw{} for SN+DNF prior. In Figure \ref{fig:deltaw_ccia}, we show the data $\Delta w$ values with an uncertainty which is given by the standard deviation of 25 $\Delta w$ values from simulations. The data $\Delta w$ values are consistent with simulations to $\sim1\sigma$ for these three redshift cases and subdominant to the overall uncertainty on $w$.

However, our data results for SN+flat prior and SN+BPZ prior are significantly discrepant with our simulation results. We are unable to attribute this to any particular cause but note that i) $\sigma_{\rm int}$ is higher for these two cases across data and simulations than the others and ii) the redshift scatter and outlier rate are higher in the simulations than the data. The largest differences between each of these \photoz\ cases are in the Hubble residuals for redshift bins at $z < 0.3$, but there are no clear trends in either the simulations or data that explain the final $\Delta w$ discrepancies. We also note that unlike in the simulations, using SN+SOMPZ $p(z)$ prior and SN+SOMPZ Gaussian prior estimates do not result in a significantly smaller final sample.

\subsection{Error scaling for Distance Modulus Uncertainties}
\label{sec:errorscaling}
In Section \ref{sec:mu_uncert} we describe the multiplicative error scaling term introduced in the original BBC framework to address naively overestimated distance uncertainties. Recent analyses such as \citet{DES5YR} and \citet{PantheonPlus} have also included an additive (in quadrature) term $\sigma_{\rm floor}^2 (z,c,M_*)$ in $\sigma_{\mu}^2$, which is intended to allow for additional scatter beyond those computed in $\sigma_{\textrm{SALT3}}$. However, we find that the inclusion of this additive term results in a BBC reduced $\chi^2$ significantly $<1$. The inclusion of the error floor results in overestimated uncertainties, which we hypothesize could be due to the fact that bias corrections are computed at the measured redshift rather than the true redshift, which may cause redshift bin migrations and therefore mismeasured error scaling factors from BBC. As we are able to recover the input cosmology with reduced $\chi^2 \sim 1$ with only the multiplicative error scaling and without the additive error floor, we leave investigation of this subtlety to future work.                 

\section{Discussion}
\label{sec:discussion}

While there are clear benefits to expanding the size of our usable SN samples using photometric redshifts for SN Ia cosmology, this approach also presents new systematics to understand and quantify. Here we have developed efforts showing that true cosmological parameters can be estimated using photometrically classified SNe with \photoz\ for a cosmological analysis. We use realistic simulated \photoz, photometric classification to address non-Ia contaminants, and a dust-based intrinsic scatter model. In this section we detail further work that is needed to rigorously measure unbiased cosmological parameters in a completely photometric analysis.

\subsection{Systematic Uncertainties}
\label{sec:syst_uncert}
In this work we have focused primarily on the recovery of unbiased cosmological parameters relative to those inferred with \specz\ by considering $\Delta w$ values. However, future cosmological analyses will require a full accounting of systematics (calibration, intrinsic scatter model, etc.) using a statistical+systematic covariance matrix \citep{Conley11} as in \cite{DES5YR}. 

More realistic redshift systematics tests and systematic variations for a covariance matrix will rely on a systematic uncertainty quantification from a given \photoz\ algorithm. One such example is the rigorously quantified sources of uncertainty from SOMPZ, which include contributions from shot noise and sample variance, inherent SOMPZ method uncertainty, redshift sample uncertainty, photometric calibration uncertainty, and synthetic source injection method uncertainty (Table 2, \citealp{Myles21}). As the uncertainties are given on the mean of each of four tomographic bins, one can interpolate across the sample redshift range to consider a redshift-dependent bias for individual host galaxies. However, other \photoz\ algorithms tend to include only statistical uncertainties with less obvious or explicit choices for systematic variations. We emphasize that it is most important to understand any redshift-dependent systematics, as they can be highly degenerate with cosmological constraints. Future cosmological analyses will require detailed studies of the modeling choices for systematics beyond a 1D analytic description of \photoz\ bias. 

While here we study the primary effect of using \photoz\ on the Hubble diagram, there are several second order systematics that are also reliant on redshifts and will require further study for an entirely photometric SN cosmology analysis. Two of the leading systematic uncertainties in the main DES-SN5YR analysis are those of SN intrinsic scatter and SN correlations with their host galaxies. In this work we relied on only one realization of a dust-based intrinsic scatter model with parameters constrained using spectroscopic redshifts. As these parameters are constrained using SN color, Hubble residuals, and Hubble scatter, using \photoz\ for the SN light-curve fit may cause biases which will need to be further studied. While here we have used host galaxy stellar masses obtained using host galaxy photometry with spectroscopic redshifts, future studies may require more detailed consideration of how galaxy mass estimates may be biased by using photo-$z$ instead, as these effects also propagate to the SN intrinsic scatter model.

\subsection{Future Work}
\label{sec:futurework}
As described in Section \ref{sec:BBC}, we use a binned Hubble diagram for this analysis, i.e. bias-corrected distances are determined in redshift bins. \cite{BHS21} and \cite{Kessler23} have shown that an unbinned Hubble diagram and covariance matrix (for spectroscopically confirmed samples) or re-binned Hubble diagram (for photometrically confirmed samples) result in a $1.5\times$ smaller systematic uncertainty compared to a binned Hubble diagram. However, previous work \citep{Mitra23} has shown that using an unbinned Hubble diagram with photometric redshifts results in smaller uncertainties but unresolved cosmological biases. This outstanding issue must be addressed in order to take advantage of the so-called data ``self-calibration'' which allows for the most robust results with a given dataset. 

While analyses with spectroscopic redshifts have treated redshift uncertainties as negligible, a crucial component that has not yet been fully developed is the treatment of redshift uncertainties when they are non-negligible, as is the case for photometric redshifts. In this analysis we discuss the use of host galaxy \photoz\ PDFs as priors in the SN+host redshift fit, but each SN is ultimately placed on the Hubble diagram with a single redshift without its (non-Gaussian) redshift uncertainties accounted for in the cosmology fit. \cite{Roberts17} presents a potential resolution to this using a hierarchical Bayesian framework inspired by the BEAMS formalism that marginalizes over redshift uncertainties to simultaneously address photometric classification, host misassociation, and photometric redshifts. 

\citet{RuhlmannKleider22} investigate two other potential methods for propagating redshift uncertainties to cosmology: i) refitting individual photometric redshifts simultaneously with cosmology and ii) sampling from the redshift probability distribution and propagating to the cosmology fit $\chi^2$ values. For a flat $\Lambda$CDM model, they find that sampling the redshift resolution function reduces bias in $\Omega_M$ to $0.3\sigma$ as long as the sample is bias corrected accounting for \photoz\ bias and core-collapse contamination. They also conclude that propagating the redshift resolution to the cosmological likelihood is likely a method of secondary importance for reducing biases compared to i) containing CC contamination to below the few percent level, ii) relying on spectroscopic redshifts at $z < 0.5$, as redshift biases lead to stronger shifts in luminosity distances at low-$z$, and iii) ensuring bias corrections include \photoz\ modeling.

As shown in Section \ref{sec:sn+hostz}, the use of the full $p(z)$ as a prior for a SN+host \photoz\ was not more informative than a prior approximated as a Gaussian. Future analyses could consider developing the cosmological framework required to use a $p(z)$ as the final redshift information on the Hubble diagram, rather than solely as a prior. This could involve computing a weighted distance contribution at various sampled redshifts from a PDF, similar to \citet{RuhlmannKleider22}, but would also likely require the development of an emulator or rapid light-curve fitting for computational feasibility in order to calculate distances at multiple redshifts for thousands of supernovae.

In this study, we have limited our sample to SNe with \specz\ available; if this requirement is lifted, a larger cosmological DES-SN sample can be built using \photoz\ only, as investigated in \citet{Moller24}. We estimate this could comprise a sample size of at least $\sim 2200$ SNe after cosmological cuts, compared to the $\sim 1600$ in the DES-SN5YR photometric sample. While here we have used all photo-$z$s available without having to rerun algorithms, an alternative approach to using \photoz\ for SN cosmology could be to subselect a sample with higher precision \photoz\ only, such as by introducing a cut on the \photoz\ uncertainty. Subsamples of SNe may also have overlapping information from many-band photometric surveys with high-quality, many-band \photoz\ for a subsample of SNe. Recent and upcoming surveys with Euclid and Roman will provide additional information in NIR bands, which can also be used to drastically improve \photoz\ estimates. The performance of other photometric redshift algorithms beyond those considered in this work should also be investigated, e.g. from \cite{Qu23b}, which uses a convolutional neural network to estimate \photoz\ PDFs from SN light-curves without host information. 

Here we have focused on a cosmological SN sample which is entirely reliant on photometric redshifts, but in the future it will likely be more realistic and ideal to use a combination of \specz\ when available and \photoz\ when not (at higher redshifts). This will require careful attention to the modeling of spectroscopic and photometric redshift efficiencies and study of the impact of the redshift cutoff at which \specz\ are available.

\section{Conclusion}
In this work, we have shown that SN+host \photoz\ are sufficient for an unbiased Type Ia Supernova cosmology analysis. Using \usableSNe{} SNe from the DES-SN5YR photometrically classified sample and fitting \photoz\ from the SN light-curve with host-galaxy priors, we obtain $\pm\Delta w$ values as low as $0.01$ and consistently subdominant to the overall $w$ uncertainty. With SN+host \photoz\ estimates using SOMPZ and DNF priors we are able to obtain $w$ estimates consistent with those obtained using \specz\ within uncertainties. We also show, using simulations, that our analysis methods are robust. We present the first efforts toward understanding host-galaxy \photoz\ modeling and algorithm related systematics in preparation for future SN Ia cosmology analyses to be done with photometric redshifts and highlight remaining open questions to be addressed in upcoming works.

\section*{Acknowledgements}

\textbf{Author contributions:} RC developed and performed the analysis and wrote the manuscript. DS, MV, ER, JM, RK, and BP assisted in developing the project and provided useful discussion and insight. MSa and MSm served as internal reviewers. All authors contributed to this paper and/or carried out infrastructure work that made this analysis possible. Additional highlighted contributions include: \textit{Construction and validation of the DES-SN5YR data and Hubble diagram}: AM, BP, BS, CL, DB, DS, GQ, JL, LG, MSa, MSm, MSu, PA, PW, MV.  \textit{Contributed to the internal review process}: AM, CL, HQ, JL, LG, MSa, MSm, MSu, PW, TD. \textit{Development and/or maintenance of software}: PA, RK.
The remaining authors have made contributions to this paper that include, but are not limited to, the construction of DECam and other aspects of collecting the data; data processing and calibration; developing broadly used methods, codes, and simulations; running the pipelines and validation tests; and promoting the science analysis.

This material is based upon work supported by the U.S. Department of Energy, Office of Science, Office of Workforce Development for Teachers and Scientists, Office of Science Graduate Student Research (SCGSR) program. The SCGSR program is administered by the Oak Ridge Institute for Science and Education for the DOE under contract number DE-SC0014664.

DS is supported by Department of Energy grant DESC0010007, the David and Lucile Packard Foundation, the Templeton Foundation and Sloan Foundation. LG acknowledges financial support from the Spanish Ministerio de Ciencia e Innovaci\'on (MCIN) and the Agencia Estatal de Investigaci\'on (AEI) 10.13039/501100011033 under the PID2020-115253GA-I00 HOSTFLOWS project, from Centro Superior de Investigaciones Cient\'ificas (CSIC) under the PIE project 20215AT016 and the program Unidad de Excelencia Mar\'ia de Maeztu CEX2020-001058-M, and from the Departament de Recerca i Universitats de la Generalitat de Catalunya through the 2021-SGR-01270 grant.

This paper has gone through internal review by the DES collaboration. 
Funding for the DES Projects has been provided by the U.S. Department of Energy, the U.S. National Science Foundation, the Ministry of Science and Education of Spain, the Science and Technology Facilities Council of the United Kingdom, the Higher Education Funding Council for England, the National Center for Supercomputing Applications at the University of Illinois at Urbana-Champaign, the Kavli Institute of Cosmological Physics at the University of Chicago, the Center for Cosmology and Astro-Particle Physics at the Ohio State University, the Mitchell Institute for Fundamental Physics and Astronomy at Texas A\&M University, Financiadora de Estudos e Projetos, Fundação Carlos Chagas Filho de Amparo à Pesquisa do Estado do Rio de Janeiro, Conselho Nacional de Desenvolvimento Científico e Tecnológico and the Ministério da Ciência, Tecnologia e Inovação, the Deutsche Forschungsgemeinschaft and the Collaborating Institutions in the Dark Energy Survey.

The Collaborating Institutions are Argonne National Laboratory, the University of California at Santa Cruz, the University of Cambridge, Centro de Investigaciones Energéticas, Medioambientales y Tecnológicas-Madrid, the University of Chicago, University College London, the DES-Brazil Consortium, the University of Edinburgh, the Eidgenössische Technische Hochschule (ETH) Zürich, Fermi National Accelerator Laboratory, the University of Illinois at Urbana-Champaign, the Institut de Ciències de l’Espai (IEEC/CSIC), the Institut de Física d’Altes Energies, Lawrence Berkeley National Laboratory, the Ludwig-Maximilians Universität München and the associated Excellence Cluster Universe, the University of Michigan, NFS’s NOIRLab, the University of Nottingham, The Ohio State University, the University of Pennsylvania, the University of Portsmouth, SLAC National Accelerator Laboratory, Stanford University, the University of Sussex, Texas A\&M University, and the OzDES Membership Consortium.

Based in part on observations at Cerro Tololo Inter-American Observatory at NSF’s NOIRLab (NOIRLab Prop. ID 2012B-0001; PI: J. Frieman), which is managed by the Association of Universities for Research in Astronomy (AURA) under a cooperative agreement with the National Science Foundation. 

Based in part on data acquired at the Anglo-Australian Telescope, under program A/2013B/012. We acknowledge the traditional owners of the land on which the AAT stands, the Gamilaraay people, and pay our respects to elders past and present.

The DES data management system is supported by the National Science Foundation under Grant Numbers AST-1138766 and AST-1536171. The DES participants from Spanish institutions are partially supported by MICINN under grants ESP2017-89838, PGC2018-094773, PGC2018-102021, SEV-2016-0588, SEV-2016-0597, and MDM-2015-0509, some of which include ERDF funds from the European Union. IFAE is partially funded by the CERCA program of the Generalitat de Catalunya. Research leading to these results has received funding from the European Research Council under the European Union’s Seventh Framework Program (FP7/2007-2013) including ERC grant agreements 240672, 291329, and 306478. We acknowledge support from the Brazilian Instituto Nacional de Ciência e Tecnologia (INCT) do e-Universo (CNPq grant 465376/2014-2).

This work was completed in part with resources provided by the University of Chicago’s Research Computing Center.

\section*{Data Availability}

Data used in this article is publicly available from the DES-SN5YR Data Release \citep{Sanchez24} on Github (``{\tt github.com/des-science/DES-SN5YR}'').

\section*{Affiliations}
$^{1}$ Department of Physics, Duke University Durham, NC 27708, USA\\
$^{2}$ NASA Einstein Fellow\\
$^{3}$ Department of Physics, University of Oxford, Denys Wilkinson Building, Keble Road, Oxford OX1 3RH, United Kingdom\\
$^{4}$ Kavli Institute for Particle Astrophysics \& Cosmology, P. O. Box 2450, Stanford University, Stanford, CA 94305, USA\\
$^{5}$ SLAC National Accelerator Laboratory, Menlo Park, CA 94025, USA\\
$^{6}$ Department of Astrophysical Sciences, Princeton University, Peyton Hall, Princeton, NJ 08544, USA\\
$^{7}$ Department of Astronomy and Astrophysics, University of Chicago, Chicago, IL 60637, USA\\
$^{8}$ Kavli Institute for Cosmological Physics, University of Chicago, Chicago, IL 60637, USA\\
$^{9}$ Univ Lyon, Univ Claude Bernard Lyon 1, CNRS, IP2I Lyon / IN2P3, IMR 5822, F-69622, Villeurbanne, France\\
$^{10}$ Department of Physics and Astronomy, University of Pennsylvania, Philadelphia, PA 19104, USA\\
$^{11}$ School of Physics and Astronomy, University of Southampton,  Southampton, SO17 1BJ, UK\\
$^{12}$ The Research School of Astronomy and Astrophysics, Australian National University, ACT 2601, Australia\\
$^{13}$ Department of Astronomy, Boston University, 725 Commonwealth Ave., Boston, MA 02215, USA\\
$^{14}$ Department of Physics, Boston University, 590 Commonwealth Ave., Boston, MA 02215, USA\\
$^{15}$ School of Mathematics and Physics, University of Queensland,  Brisbane, QLD 4072, Australia\\
$^{16}$ Institut d'Estudis Espacials de Catalunya (IEEC), 08034 Barcelona, Spain\\
$^{17}$ Institute of Space Sciences (ICE, CSIC),  Campus UAB, Carrer de Can Magrans, s/n,  08193 Barcelona, Spain\\
$^{18}$ Centre for Gravitational Astrophysics, College of Science, The Australian National University, ACT 2601, Australia\\
$^{19}$ Centre for Astrophysics \& Supercomputing, Swinburne University of Technology, Victoria 3122, Australia\\
$^{20}$ Aix Marseille Univ, CNRS/IN2P3, CPPM, Marseille, France\\
$^{21}$ Cerro Tololo Inter-American Observatory, NSF's National Optical-Infrared Astronomy Research Laboratory, Casilla 603, La Serena, Chile\\
$^{22}$ Laborat\'orio Interinstitucional de e-Astronomia - LIneA, Rua Gal. Jos\'e Cristino 77, Rio de Janeiro, RJ - 20921-400, Brazil\\
$^{23}$ Fermi National Accelerator Laboratory, P. O. Box 500, Batavia, IL 60510, USA\\
$^{24}$ Department of Physics, University of Michigan, Ann Arbor, MI 48109, USA\\
$^{25}$ Physik-Institute, University of Zurich, Winterthurerstrasse 190, 8057 Zurich, Switzerland\\
$^{26}$ Institute of Cosmology and Gravitation, University of Portsmouth, Portsmouth, PO1 3FX, UK\\
$^{27}$ Department of Physics \& Astronomy, University College London, Gower Street, London, WC1E 6BT, UK\\
$^{28}$ Instituto de Astrofisica de Canarias, E-38205 La Laguna, Tenerife, Spain\\
$^{29}$ Institut de F\'{\i}sica d'Altes Energies (IFAE), The Barcelona Institute of Science and Technology, Campus UAB, 08193 Bellaterra (Barcelona) Spain\\
$^{30}$ NASA Goddard Space Flight Center, 8800 Greenbelt Rd, Greenbelt, MD 20771, USA\\
$^{31}$ Jodrell Bank Center for Astrophysics, School of Physics and Astronomy, University of Manchester, Oxford Road, Manchester, M13 9PL, UK\\
$^{32}$ University of Nottingham, School of Physics and Astronomy, Nottingham NG7 2RD, UK\\
$^{33}$ Hamburger Sternwarte, Universit\"{a}t Hamburg, Gojenbergsweg 112, 21029 Hamburg, Germany\\
$^{34}$ Jet Propulsion Laboratory, California Institute of Technology, 4800 Oak Grove Dr., Pasadena, CA 91109, USA\\
$^{35}$ Institute of Theoretical Astrophysics, University of Oslo. P.O. Box 1029 Blindern, NO-0315 Oslo, Norway\\
$^{36}$ Instituto de Fisica Teorica UAM/CSIC, Universidad Autonoma de Madrid, 28049 Madrid, Spain\\
$^{37}$ University Observatory, Faculty of Physics, Ludwig-Maximilians-Universit\"at, Scheinerstr. 1, 81679 Munich, Germany\\
$^{38}$ Center for Astrophysical Surveys, National Center for Supercomputing Applications, 1205 West Clark St., Urbana, IL 61801, USA\\
$^{39}$ Department of Astronomy, University of Illinois at Urbana-Champaign, 1002 W. Green Street, Urbana, IL 61801, USA\\
$^{40}$ Santa Cruz Institute for Particle Physics, Santa Cruz, CA 95064, USA\\
$^{41}$ Center for Cosmology and Astro-Particle Physics, The Ohio State University, Columbus, OH 43210, USA\\
$^{42}$ Department of Physics, The Ohio State University, Columbus, OH 43210, USA\\
$^{43}$ Center for Astrophysics $\vert$ Harvard \& Smithsonian, 60 Garden Street, Cambridge, MA 02138, USA\\
$^{44}$ Australian Astronomical Optics, Macquarie University, North Ryde, NSW 2113, Australia\\
$^{45}$ Lowell Observatory, 1400 Mars Hill Rd, Flagstaff, AZ 86001, USA\\
$^{46}$ Departamento de F\'isica Matem\'atica, Instituto de F\'isica, Universidade de S\~ao Paulo, CP 66318, S\~ao Paulo, SP, 05314-970, Brazil\\
$^{47}$ George P. and Cynthia Woods Mitchell Institute for Fundamental Physics and Astronomy, and Department of Physics and Astronomy, Texas A\&M University, College Station, TX 77843,  USA\\
$^{48}$ LPSC Grenoble - 53, Avenue des Martyrs 38026 Grenoble, France\\
$^{49}$ Instituci\'o Catalana de Recerca i Estudis Avan\c{c}ats, E-08010 Barcelona, Spain\\
$^{50}$ Observat\'orio Nacional, Rua Gal. Jos\'e Cristino 77, Rio de Janeiro, RJ - 20921-400, Brazil\\
$^{51}$ Department of Physics, Carnegie Mellon University, Pittsburgh, Pennsylvania 15312, USA\\
$^{52}$ Department of Physics, Northeastern University, Boston, MA 02115, USA\\
$^{53}$ Centro de Investigaciones Energ\'eticas, Medioambientales y Tecnol\'ogicas (CIEMAT), Madrid, Spain\\
$^{54}$ Computer Science and Mathematics Division, Oak Ridge National Laboratory, Oak Ridge, TN 37831\\
$^{55}$ Department of Astronomy, University of California, Berkeley,  501 Campbell Hall, Berkeley, CA 94720, USA\\
$^{56}$ Lawrence Berkeley National Laboratory, 1 Cyclotron Road, Berkeley, CA 94720, USA\\
$^{57}$ Max Planck Institute for Extraterrestrial Physics, Giessenbachstrasse, 85748 Garching, Germany\\
$^{58}$ Universit\"ats-Sternwarte, Fakult\"at f\"ur Physik, Ludwig-Maximilians Universit\"at M\"unchen, Scheinerstr. 1, 81679 M\"unchen, Germany\\



\bibliographystyle{mnras}
\bibliography{research2} 




\appendix

\section{Simulation and data comparisons for photo-$z$}
Following Section \ref{sec:datasimcomp} we show comparison plots of light-curve parameters between the simulations and data when redshift is fit jointly using different host-galaxy \photoz\ priors.

\begin{figure*}
	\centering
    \includegraphics[width=\textwidth]{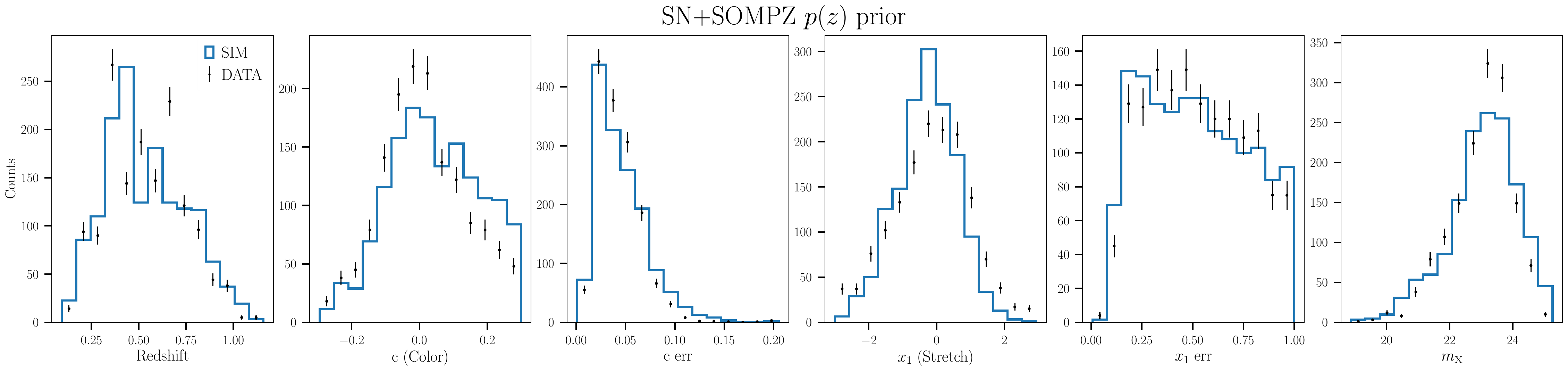}
    \includegraphics[width=\textwidth]{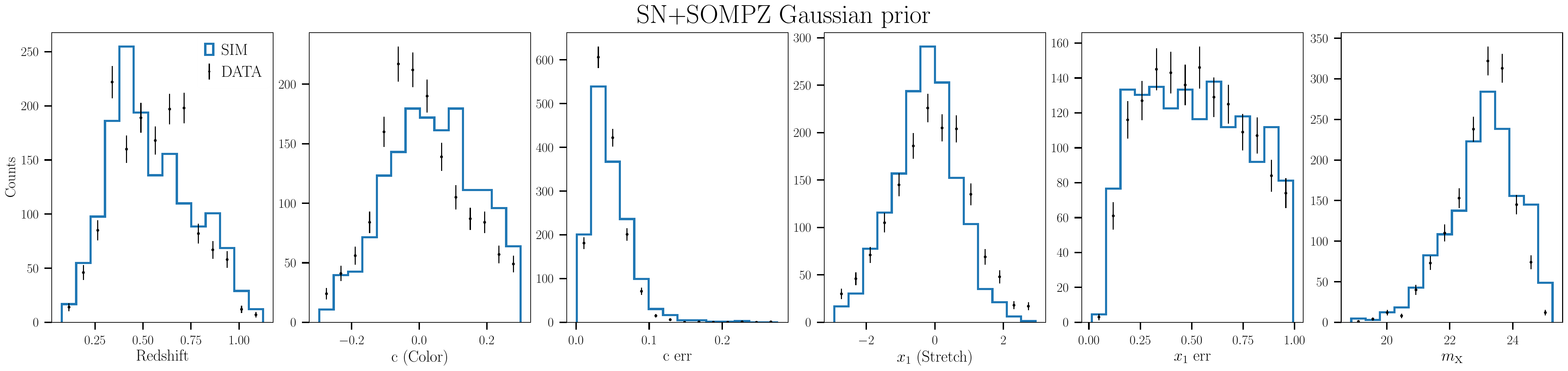}
    \includegraphics[width=\textwidth]{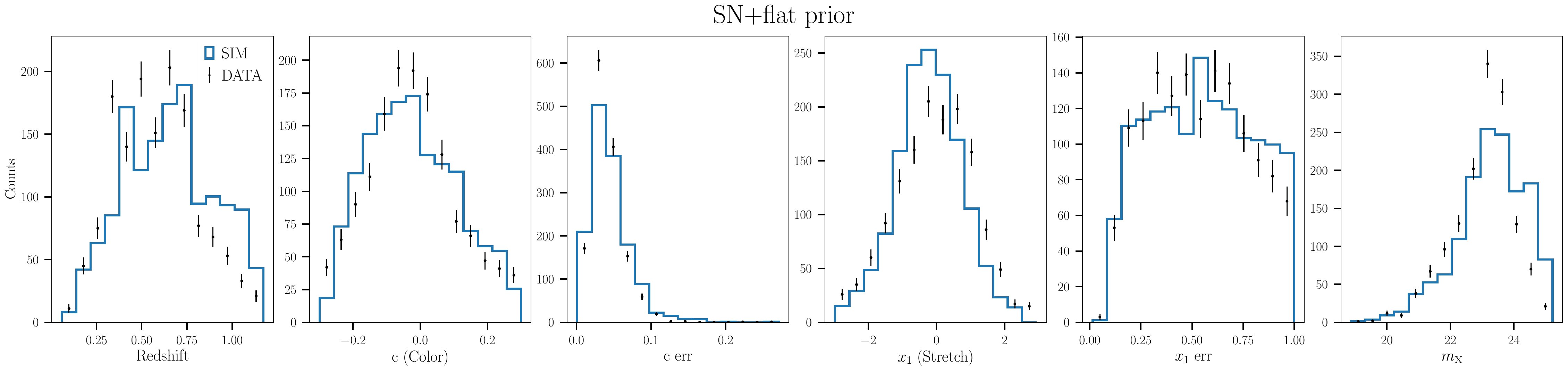}
    \includegraphics[width=\textwidth]{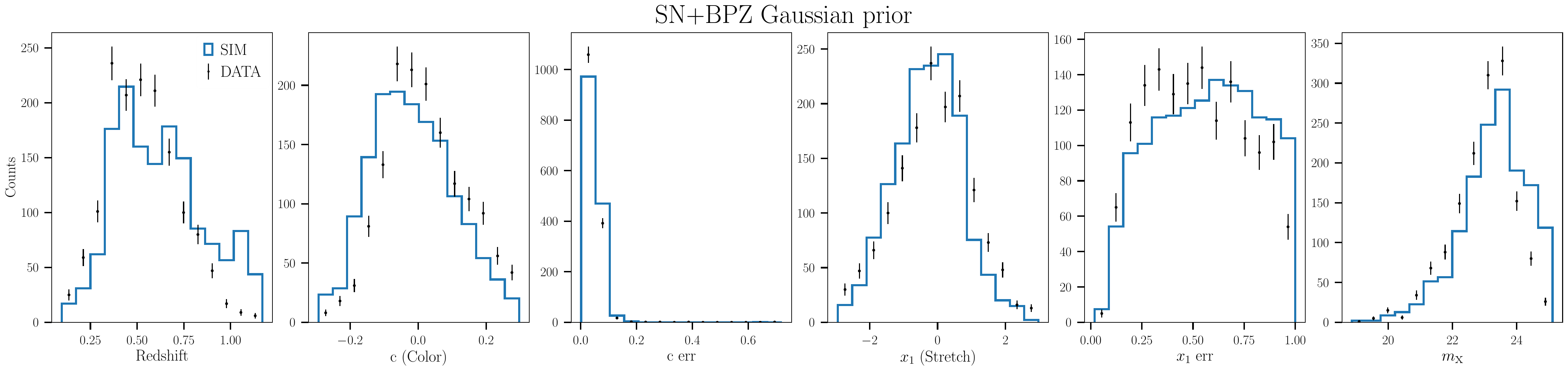}
    \includegraphics[width=\textwidth]{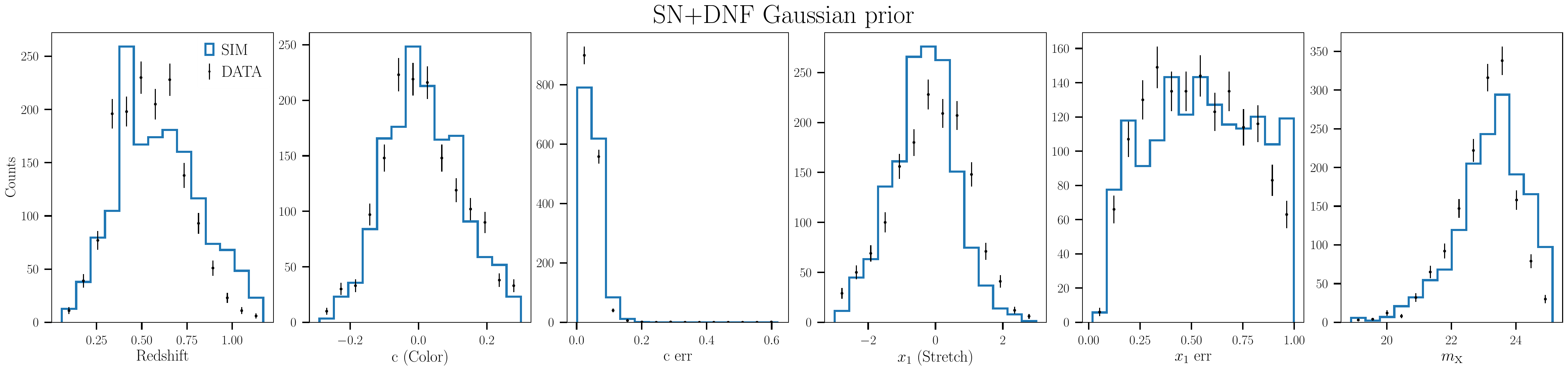}
    \caption{Comparison of fitted SN light-curve parameters and their errors, as in Fig. \ref{fig:sim_data}.}
\end{figure*}

\section{Hubble diagrams}
Here we present the Hubble diagrams for each \photoz\ case for simulations and data as detailed in Section \ref{sec:results}.

\begin{figure*}
	\centering
    \includegraphics[width=\textwidth]{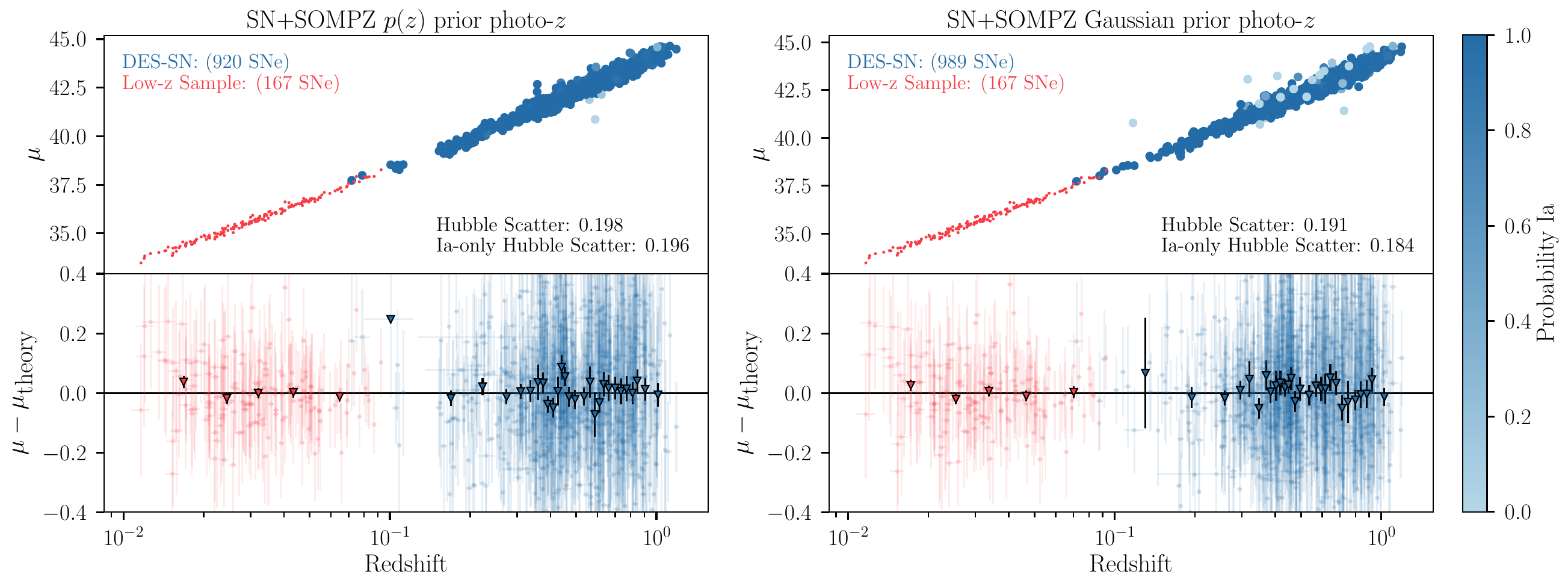}
    \includegraphics[width=\textwidth]{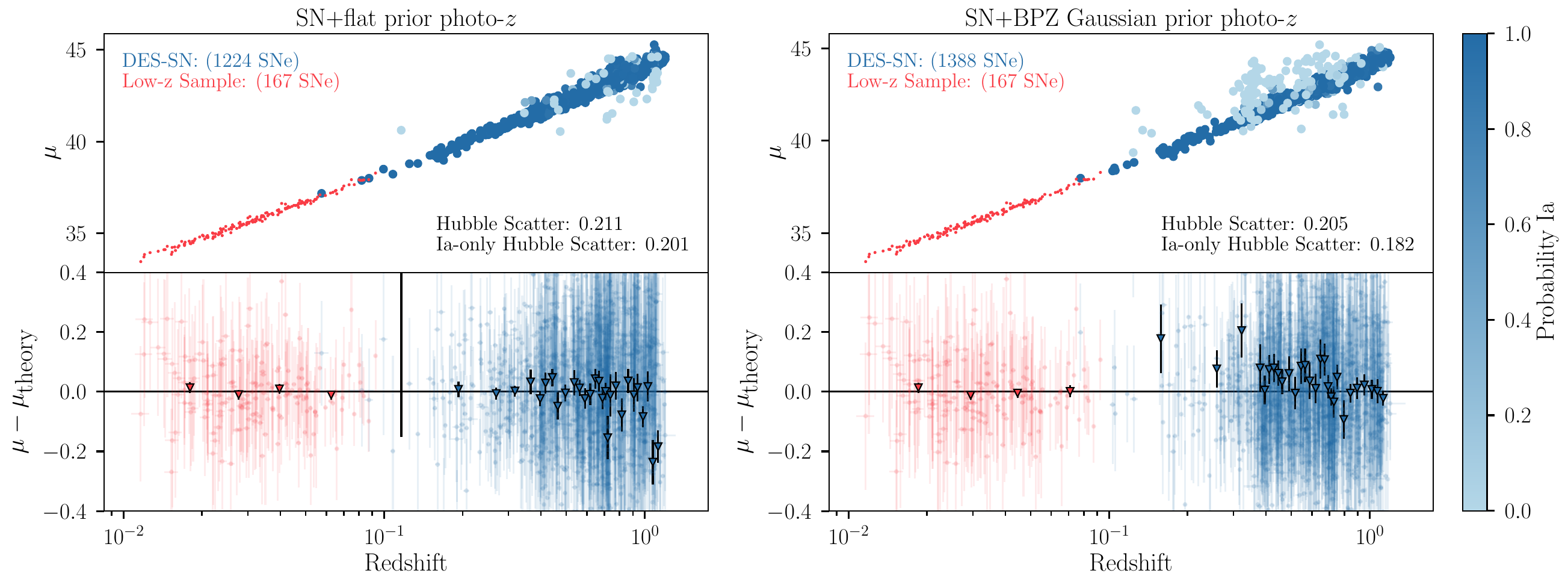}
	\caption{Simulation Hubble diagrams for each redshift variant. Each point is shaded by the probability of being a Ia as given by SNN.}
	\label{fig:simHD}
\end{figure*}

\begin{figure*}
	\centering
    \includegraphics[width=\textwidth]{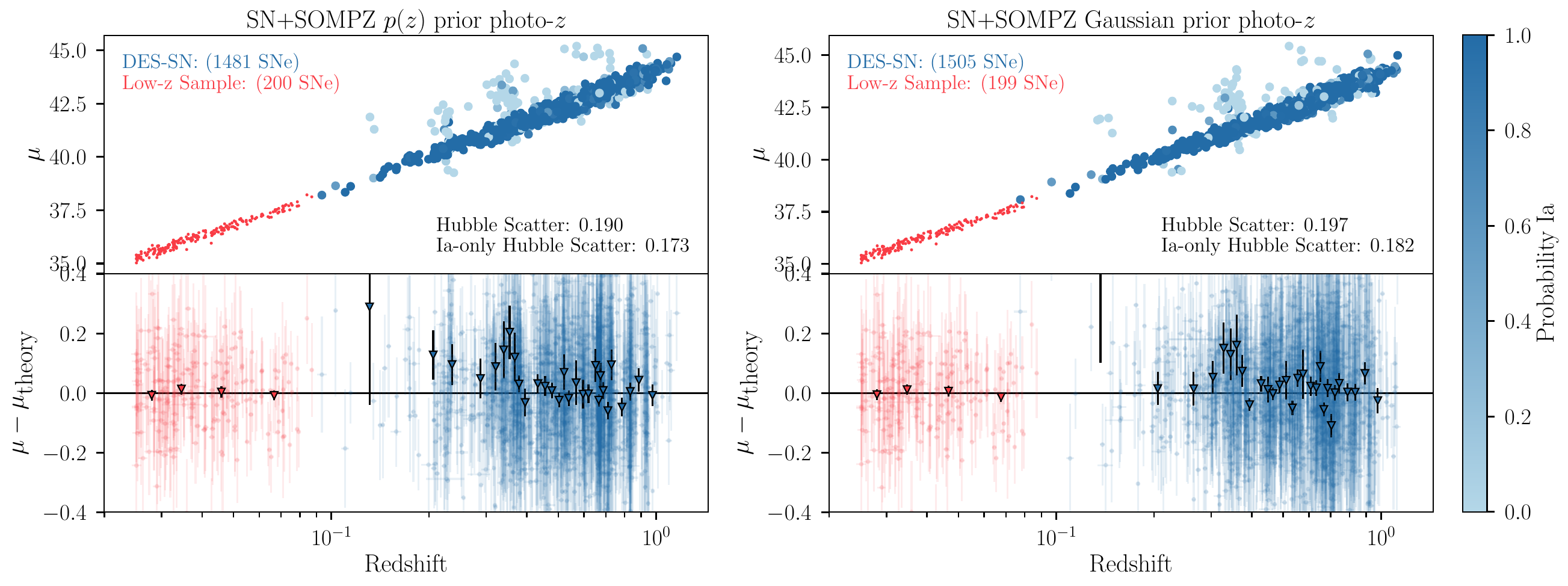}
    \includegraphics[width=\textwidth]{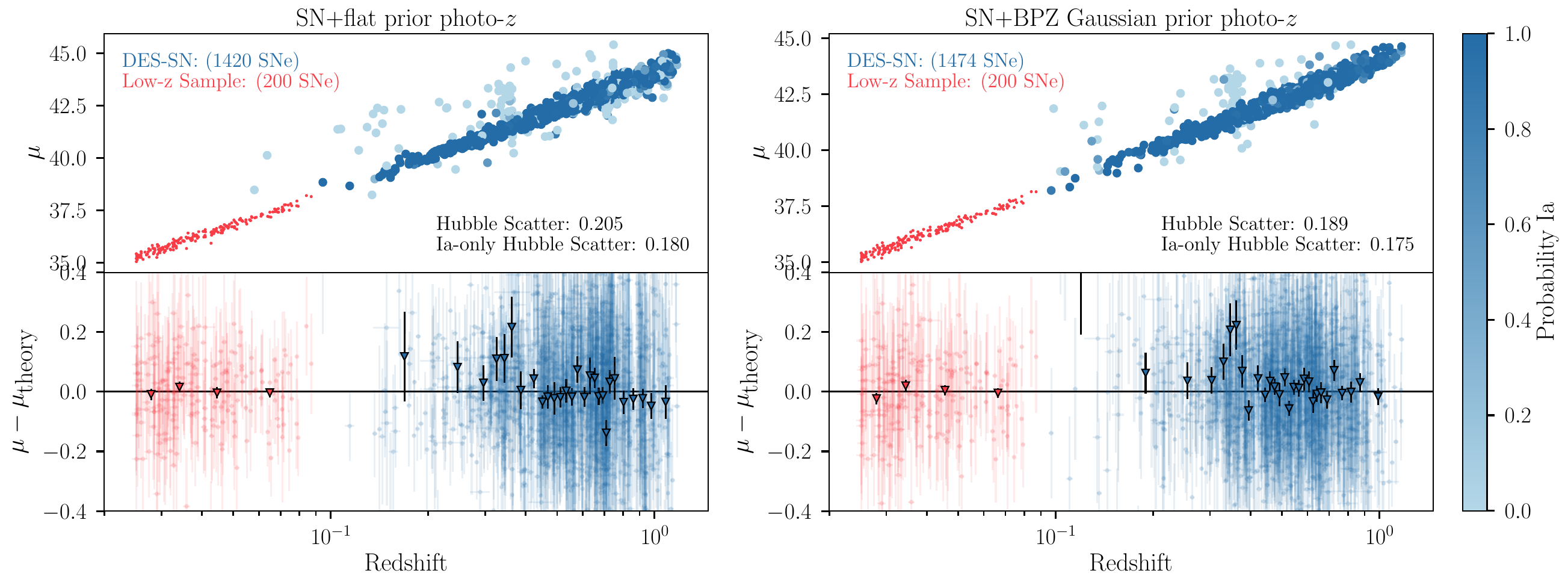}
	\caption{Data Hubble diagrams for each redshift variant. Each point is shaded by the probability of being a Ia as given by SNN.}
	\label{fig:dataHD}
\end{figure*}


\bsp	
\label{lastpage}
\end{document}